\documentclass[12pt]{article}
\usepackage{epsfig}
\usepackage{amsfonts}
\begin {document}

\title {ON THE EXISTENCE OF MSA COORDINATES}

\author{J.L. Hern\'andez-Pastora\thanks{E.T.S. Ingenier\'\i a
Industrial de B\'ejar. Phone: +34 923 408080 Ext
2263. Also at +34 923 294400 Ext 1527. e-mail address: jlhp@usal.es}\\
\\
Departamento de Matem\'atica Aplicada. \\ Universidad de Salamanca.  Salamanca,
Espa\~na.  }

\date{} \maketitle

\begin{abstract}
The static solutions of the axially symmetric vacuum Einstein equations with a finite number of Relativistic Multipole Moments (RMM) are described by means of a function that can be written in the same analytic form as the  Newtonian gravitational multipole  potential. A family of so-called MSA (Multipole-Symmetry Adapted) coordinates  are introduced and calculated at any multipole order to perform the transformation of the Weyl solutions.

In analogy with a previous result \cite{cqg} obtained in Newtonian gravity, the existence of a symmetry of a  certain system of differential equations leading to the determination of that kind of multipole solutions in General Relativity is explored. The relationship between the existence of this kind of coordinate and the  symmetries mentioned is proved for some cases, and the characterization of the MSA system of coordinates by means of this relationship is discussed.

\end{abstract}


\newpage

\section{Introduction}

As is known, the description of Newtonian Gravity (NG) in the vacuum involves solutions of the Laplace equation whose general well-behaved solution is a series with arbitrary constants that can be identified with the Multipole Moments (MM) of the source, and these quantities allow us to characterize the specific solutions given by the succession of partial sums of the series.

In contrast,  the static solutions of the axially symmetric Eisntein vacuum equations describing the gravitational field of a bounded isolated mass distribution in General Relativity (GR) can be described by means of only one metric function, $f\equiv g_{00}$, which satisfies the Ernst equation \cite{ernst}.

We are interested in the following questions: might it  be possible to obtain a description of these
solutions in GR by means of a function, namely $u$,  with the same behaviour as the classical potential? Could we write this function $u$ analytically equal to the Newtonian gravitational series but in terms of  Relativistic Multipole Moments (RMM)?

We are concerned with these questions for several reasons, in particular because such a description of the relativistic gravitational solution would recover the benefits of the classical interpretation of the gravitational potential (see \cite{cqg} for details). Moreover, the Weyl family of solutions depends on arbitrary constants, $a_n$,  in principle without any physical criteria to choose one or another solution from them, whereas the function $u$ would allow us to deal, in a very simple form, with the Relativistic Multipole Solutions. This has been the aim of some authors and their works devoted to obtaining solutions of the Einstein vacuum equations with a finite number of prescribed RMM.

In this work we seek an answer to these questions by introducing a family of coordinate systems referred to as  MSA (Multipole-Symmetry Adapted). The possibility of  extrapolating  the symmetries obtained in NG \cite{cqg} to GR, as well as  characterizing the solutions with a finite number of RMM by means of group-invariant solutions, are the relevant features of these coordinate systems and the reason for their proposed name.

In a work published recently  \cite{cqg}, the existence of some kinds of  symmetries in NG has been proved, which  makes it possible to extract from all solutions of the axially symmetric Laplace equation those with the prescribed Newtonian Multipole Moments.
A family of vector fields that  are the infinitesimal generators of certain one-parameter groups of transformations can be constructed. These vector fields represent symmetries of certain  systems of differential equations whose
group-invariant solutions   turn out  to be the family of axisymmetric potentials related to specific gravitational multipoles.

By introducing these coordinates, the function $u$, which describes the static and axially symmetric vacuum solution with a finite number of RMM, should satisfy the same system of differential equations as the classical potential in NG, and  the symmetries of these equations \cite{cqg} thus allow us to describe and determine the Multipole Solutions in GR analogously to the Newtonian case.

Since the function $u$, which is  transformed from $g_{00}$, must fulfil the corresponding Ernst equation written in MSA coordinates, the following question arises: is it possible to  obtain conditions on the change of coordinates, to choose the suitable gauge, from the  extension of  the symmetries to the corresponding Ernst equation? In other words, can the symmetry groups obtained for the Laplace equation and the supplementary equation \cite{cqg} be extrapolated to the Ernst equation in that system of coordinates? And if  so, could we establish theorems relating the existence of the symmetry of a system of differential equations to that system of coordinates?  We shall see, at least for the Monopole case, that this relationship can be used to determine  the MSA radial coordinate explicitly.

We shall try to answer  these questions along the work in the  following way:

In section 2, the MSA systems of coordinates are defined from the context of the multipole expansion of gravitation introduced by Thorne \cite{thorne}, and the procedure to calculate these coordinates is shown  for each set of multipole structure of the desired solution. The procedure first consists of performing a  coordinate transformation from Weyl coordinates, preserving the Killing vectors and the asymptotically Cartesian behaviour. Second, we introduce a function $u$ by redefining the $g_{00}$ metric component and we force this function $u$  to be a solution of the corresponding Ernst equation written in the new system of coordinates. The results are addressed in Appendix B. Some comments about the behaviour and interpretation of the coordinates obtained complete this section.

In section 3 we attempt to provide these coordinates with a meaningful interpretation by means of the existence of symmetries of certain differential equations. We recall that in these coordinates the static and axially symmetric vacuum solutions with a finite number of RMM can be described as group-invariant solutions of the same system of differential equations that admits the symmetries obtained for the Newtonian case \cite{cqg}.  Furthermore, we prove two theorems that extend those symmetries  to the corresponding Ernst equation written in the MSA system of coordinates for the Monopole and the Monopole-Dipole cases. And these theorems provide a relationship between the existence of these symmetries and the Newtonian-type solutions in GR, at least for these two cases. In subsection 3.2 a possible characterization of these systems of coordinates for each multipolar solution is explored, without taking into account our knowledge of the corresponding set of constants $a_n$ for each case. In this sense, for the Monopole case the corresponding   MSA coordinates can be obtained as the unique solution of the Ernst equation (and the suitable constraints) with  some boundary condition. Nevertheless, for any other case this procedure fails to provide the uniqueness of the MSA system, since two coordinates must be solved rather than the radial coordinate alone, as is the case only for  Spherical symmetry.
Finally, in  Appendix A the expressions of the RMM in terms of the $a_n$ constants and the inverse relation are shown for a solution with a set of arbitrary RMM up to order $10$.

\section{The MSA system of coordinates}

\subsection{Definition}

In 1980 Thorne introduced a system of coordinates called ACMC ({\it Asymptotically Cartesian and Mass Centered}) in the context of Multipole expansions of gravitational radiation \cite{thorne}. His work presents a definition of Relativistic Multipole Moments (RMM) and shows us how to deduce the RMM of a source from the form of its stationary and asymptotically flat vacuum metric in an ACMC coordinate system.

If the components of the metric are written in the coordinates $\{\hat t, r,\hat\theta, \hat\varphi\}$ we can read off the RMM from the resulting expressions,  other terms,  $R_{ij}^{(n-1)}(y)$,  called {\it Thorne rests}, appearing at the same time that are functions depending on  the angular variable $y\equiv \cos\hat\theta$ in  at least one degree lower than those associated with the RMM. For the case of axial symmetry, the $g_{00}$ component of any static metric written in that kind of coordinates resembles:
\begin{equation}
g_{00} =-1+\frac2{c^2} \left[ \sum_{n=0}^\infty \frac1{r^{n+1}}
M_{n} P_{n}(y)+ \sum_{n=1}^\infty \frac1{r^{n+1}}
R_{00}^{(n-1)}(y) \right] , \label{g00}
\end{equation}
$M_{n}$ being the RMM of order $n$, and $P_n(\omega)$ the Legendre polynomial.

There is  gauge freedom in the choice of ACMC coordinates preserving the invariance of the first series in (\ref{g00}) and addressing the differences in the metric expansion through the Thorne rests. Among the broad class of coordinate systems of this type, we wish that system to lead to an expansion of the metric in such a way that all the $R_{00}^{(n-1)}(y)$ Thorne rests will vanish. We propose that this system of coordinates should be referred to as ACMC-TRF ({\it ACMC-Thorne Rest-Free}) in a first step, and then become the so-called  MSA ({\it Multipole-Symmetry Adapted}) system of coordinates for reasons we shall see in the next section.

Thorne showed that de Donder coordinates are ACMC-to order N for all N, and  in \cite{thorne} he discussed  that a de Donder transformation of the coordinates is not necessary. Hence, we do not impose the harmonic condition on our system of coordinates, and we perform a coordinate transformation of the following form
\begin{equation}
 {\hat x}^{\alpha}=x^{\alpha}+\chi^{\alpha}(x^i) \qquad ; \qquad \chi^{\alpha}=cte+O(1/R) \ ,
\label{gauge}
\end{equation}
$\{x^{\alpha}\}$ being the Weyl spherical coordinates $\{t,R,\omega\equiv \cos\theta, \varphi\}$, and $\{\hat{x}^{\alpha}\}=\{t,r,y\equiv \cos\hat\theta,\varphi\}$ being the new system of coordinates. This keeps the Killing vectors unchanged and maintains the asymptotically flat form of the metric: $g_{\alpha \beta}=\eta_{\alpha \beta}+O(1/R)$. Now, the $\chi^{\alpha}$ functions must be adjusted so that  the metric in the new system of coordinates will satisfy the following condition: ($c=1$)
\begin{equation}
 g_{00}\equiv -1+2 u =-1+2 \sum_{n=0}^N\frac{1}{r^{n+1}} M_n P_n(y) \ , \label{u}
\end{equation}
where $N$ stands for the number of RMM we wish to consider for our metric. Henceforth, the $g_{00}$ component of the metric in this system of coordinates acquires a form that is analytically related to the Newtonian gravitational potential of a classical Multipole Solution, and (\ref{u})  represents the solution,  with a finite number (N+1) of RMM, of the static and axially symmetric Einstein vacuum equations.

\subsection{Calculation of MSA coordinates for any Multipole Solution}

The line element of a static vacuum metric is as follows:
\begin{equation}
 ds^2= -e^{2\Psi} dt^2+e^{-2\Psi+2\gamma}\left(dR^2+\frac{R^2}{1-\omega^2} 
d\omega^2\right)+e^{-2\Psi}R^2(1-\omega^2)d\varphi^2 \ ,
\label{ds2}
\end{equation}
$\{R,\omega\equiv\cos\theta,\varphi\}$ being the  Weyl spherical coordinates, and  $\Psi$, $\gamma$ are metric functions satisfying the following equations\footnote{Let us note that the integrability condition of the equation for $\gamma$ is simply the equation for $\Psi$, and therefore the solution of the Laplace equation  univocally identifies the space-time.}
\begin{eqnarray}
\triangle \Psi&\equiv& R^2\Psi_{RR}+2R\Psi_R+(1-\omega^2)\Psi_{\omega \omega}-2\omega \Psi_{\omega}=0\nonumber\\
\gamma_R &=&\frac{(1-\omega^2)}{R}\left[R^2\Psi_R^2-(1-\omega^2)\Psi_{\omega}^2-2R\omega\Psi_R\Psi_{\omega}\right]\nonumber\\
\gamma_{\omega}&=&\omega\left[R^2\Psi_R^2-(1-\omega^2)\Psi_{\omega}^2\right]+2R(1-\omega^2)\Psi_R\Psi_{\omega}\ , 
\label{einstein}
\end{eqnarray}
where the sub-indices denote partial derivation with respect to them.

The general solution for an isolated source with axial symmetry is given by the following asymptotically flat series (the  family of Weyl solutions):
\begin{equation}
 \Psi=\sum_{n=0}^{\infty}\frac{a_n}{R^{n+1}} P_n(\omega) \ , \label{psi}
\end{equation}
where coefficients $a_n$ are arbitrary constants; any set of those coefficients  univocally determines the solution.
From the calculation\footnote{We perform the FHP method \cite{fhp}, which allows us to obtain the RMM in terms of coefficients $m_n$ involved in the expansion series on the symmetry axis of a conformal Ernst potential. Since that conformal potential is related to the metric function $\Psi$ (\ref{psi}),  the coefficients $m_n$ can be expressed in terms of the set of coefficients $\{a_n\}$, and hence  the final result provides a relation $M_n=M_n(a_n)$ (\ref{A1}); the triangular structure of this relation \cite{algomio}, \cite{mio2} allows us to calculate the inverse relation, $a_n=a_n(M_n)$ (\ref{A2}).} of the RMM of this metric (\ref{ds2}) with the function $\Psi$ (\ref{psi}), one can obtain an expression for the coefficients $a_n$ in terms of the RMM and hence it is possible to  choose of these coefficients by neglecting the undesirable RMM. This procedure affords  a Multipole Solution having a finite number of Multipole Moments \cite{mio2}. Some authors have devoted some time to seeking such those solutions  (Pure Multipole Solutions in GR) \cite{mio2}, \cite{mio3}, \cite{mio4}. In \cite{mio2}, the M-Q Solution is obtained, as well as the Quadrupole Solution itself, and more recently in \cite{sueco1}, \cite{sueco2} a method has been proposed  for obtaining the general terms of the series (the coefficients $a_n$) that define  the Pure $2^N$-pole Solutions. In fact, the general term of the series corresponding to the gravitational Dipole and the solutions with Monopole plus any other $2^N$-pole moment are written specifically.

In  Appendix A, the RMM of the general static solution  (\ref{psi}) are written in terms of the coefficients $a_n$, as well as the inverse relation for a solution with arbitrary RMM up to order $10$.

Let us  set $\displaystyle{u \equiv \frac 12 (1+g_{00})}$; this function $u$, corresponding to the solution (\ref{psi}) written in Weyl coordinates ($u_{W}$), resembles the following expression
\begin{equation}
 u_W= \displaystyle{\frac 12
\left[1-exp\left(2\sum_{n=0}^{\infty}\frac{a_n}{R^{n+1}}P_n(\omega)\right)\right]}\ ,
\end{equation}
and   would provide the gravitational solution with a finite number ($N+1$) of RMM by substituting the coefficients $a_n$ from  expression (\ref{A2}).

In terms of the Ernst potential \cite{ernst}, which is a real function for the static case,  the $g_{00}$ metric component equals this potential,  and hence the function $u$ satisfies the following  equation derived from the Ernst equation \cite{ernst}:
\begin{equation}
 (2u-1)\big[R^2u_{RR}+2Ru_R+(1-\omega^2)u_{\omega \omega}-2 \omega
u_{\omega}\big]=2\big[R^2u_R^2+(1-\omega^2)u_{\omega}^2\big] \ .\label{erny2}
\end{equation}

It is straightforward to calculate the transformation of equation (\ref{erny2}) by means of an arbitrary change of coordinates from the Weyl system ($\{R,\omega\}$) to another one ($\{r,y\}$), leading to the following expressions:
\begin{eqnarray}
A u_{rr}+B u_{yy}+2 Cu_{ry}+D u_r+E u_y-\frac{2}{2u-1}\big[A u_r^2+B u_y^2+2 C u_r u_y\big]=0,
\label{ernymsa}
\end{eqnarray}
where $A$, $B$, $C$, $D$ and $E$ are functions of the coordinates $\{r,y\}$ defined as follows
\begin{eqnarray}
 A(r,y)\equiv \hat{LB}_1(r,r) \qquad B(r,y)\equiv \hat{LB}_1(y,y) \qquad 
C(r,y)\equiv\hat{LB}_1(r,y)\nonumber\\
D(r,y)\equiv \hat{LB}_2(r) \qquad E(r,y)\equiv \hat{LB}_2(y),
\label{lb1}
\end{eqnarray}
$\hat{LB}_1()$ and $\hat{LB}_2()$ being the Laplace-Beltrami operators with respect to  3-dimensional Euclidean metric (with axial symmetry) written in Weyl spherical coordinates; i.e.,
\begin{eqnarray}
 \hat{LB}_1(,) &\equiv& \eta^{ij}\nabla_i() \nabla_j()=R^2\partial_R() 
\partial_R()+(1-\omega^2)\partial_{\omega}() \partial_{\omega}()
\nonumber\\
\hat{LB}_2()&\equiv& \eta^{ij}
\nabla_{ij}()=R^2\partial^2_{RR}+2R\partial_R+(1-\omega^2)\partial^2_{\omega\omega}-2\omega\partial_{\omega}.
\label{lb2}
\end{eqnarray}

Let us perform the above-mentioned coordinate transformation (\ref{gauge}) by assuming the following asymptotically Cartesian behaviour of the new coordinates:
\begin{eqnarray}
 r=R\left[1+\sum_{n=1}^{\infty}f_n(\omega)\frac{1}{R^n}\right] \nonumber \\
y=w+\sum_{n=1}^{\infty}g_n(\omega)\frac{1}{R^n} \ .
\label{trans}
\end{eqnarray}
We shall now impose the following two conditions:
\begin{eqnarray}
(E0) \qquad & &u_W =u_{MSA}\nonumber\\
(EI) \qquad & &\big[A u_{rr}+B u_{yy}+2 Cu_{ry}+D u_r+E u_y\big]_{u=u_{MSA}}=\nonumber\\
& & \frac{2}{2u-1}\big[A u_r^2+B u_y^2+2 C u_r u_y\big]_{u=u_{MSA}}\ .
\end{eqnarray}

The equation (E0) means that we force the solution with a finite number ($N+1$) of RMM written in Weyl coordinates ($u_W$) to be functionally equal to the Newtonian gravitational potential, with that number of Multipole Moments, written in the MSA system of coordinates ($u_{MSA}$) required to exist. This condition is equivalent to the coordinate transformation of the metric component $g_{00}$.
The equation (EI) should be understood in the following way: the function $u_{MSA}$ must be a solution of the differential equation (\ref{ernymsa}) obtained from the Ernst equation for $u$ (\ref{erny2}) by means of the gauge transformation.

These two conditions  univocally determine the functions $f_n(\omega)$ and $g_n(\omega)$ (\ref{trans}), up to any order, in the following way. First, equation (E0) provides a relation between each $f_n(\omega)$ and $g_k(\omega)$, for $k$ from $1$ to $n-1$,  by developing a power series expansion on the inverse of the radial coordinate $R$.
Second, we substitute $u_{MSA}$ in the differential equation (EI) and  the resulting expression  can be expanded in power series of $1/R$ by using the gauge (\ref{trans}). Since we have considered  the previously obtained relations between functions $f_n(\omega)$ and $g_n(\omega)$ in this expansion, the condition (EI) finally leads to a complete determination of the coordinates $\{r,y\}$.

We have computed the calculation of the gauge (\ref{trans}) for a solution with the  first three RMM ($M_0$,$M_1$ and $M_2$, monopole, dipole and quadrupole moments respectively) up to order $O(1/R^{10})$; the  results obtained are shown in  Appendix B.

Let us offer some comments about the good behaviour of these coordinates $\{r,y\}$. According to the meaning of the RMM, these coordinates reveal the loss of relevance of  high-order multipoles in the description  of the solutions at large distances from the source. Moreover, as can be seen in the expressions of the functions $g_n(\omega)$ (\ref{ges}), all these functions vanish for $\omega=\pm 1$. This means that the coordinate $y$ preserves the axial symmetry, since $y=\omega$ along the axis. They are not harmonic coordinates because that condition (the de Donder gauge) is not fulfilled by the associated Cartesian coordinate $z$.\footnote{Let $\{x^i\}\equiv\{\hat x,\hat y,\hat z\}$ be Cartesian coordinates associated with the spherical ones $\{\hat r,\hat\theta,\hat\varphi\}$ as $\hat z=\hat r \cos\hat\theta$, $\hat x+i \hat y =\hat r\sin\hat\theta e^{i\hat \varphi}$; these coordinates are said to be harmonic if $\Box\{\hat x^i\}=0$, where $\Box$ denotes the D'Alambert operator with respect to the metric considered. This condition leads to the following equation for the coordinate $z$: $\hat{LB}_2(\hat z)=0$. It is easy to see that $\hat{LB}_2(\hat z)=\cos\hat \theta \hat{LB}_2(\hat r)+\hat r \hat{LB}_2(\cos\hat\theta)+2\hat {LB}_1(\hat r,\cos\hat\theta)$ and the coordinates $\{\hat r=r,\cos \hat\theta=y\}$ given in (\ref{trans}) (\ref{B7}-\ref{ges}) do not satisfy this harmonic condition.}
In addition, if we consider the MSA coordinates for the case of spherical symmetry, which can be done by neglecting all RMM\footnote{As can be seen, expressions (\ref{B1}) are recovered from (\ref{B7}) by taking all RMM, except for $M_0$, equal to zero.} in (\ref{B1}) greater than the monopole ($M_0$), then we observe that all the functions $f_n(\omega=\pm1)=0$ for $n\geq 2$, (i.e., the coordinate $r$ along the axis orthogonal to the equatorial plane shows a good behaviour; in other words, it equals the Cartesian coordinate $z$, up to a displacement along the axis). In fact, the case of spherical symmetry deserves a more detailed analysis: the expressions (\ref{B7}) for this case can be written as follows:
\begin{eqnarray}
f_{2n}(\omega)&=& M^{2n} C_{2n}^{(-1/2)}(\omega) \ , \quad f_{2n+1}(\omega)=0 \ , n\geq1, \quad f_1(\omega)=M\nonumber\\
g_{2n}(\omega)&=& -M^{2n} C_{2n+1}^{(-1/2)}(\omega) \ , \quad g_{2n+1}(\omega)=0 \ ,n\geq1, \quad g_1(\omega)=0,
\label{fqsM}
\end{eqnarray}
$C_{n}^{(-1/2)}$ being Gegenbauer orthogonal polynomials, and henceforth $M\equiv M_0$. Since the generator function of these polynomials is known:
\begin{equation}
r_{\pm}\equiv\sqrt{1\pm 2 \omega\lambda+\lambda^2}=\sum_{n=0}^{\infty}C_n^{(-1/2)}(\omega)(\mp\lambda)^n \ ,
\label{rpm}
\end{equation}
the following relations hold
\begin{eqnarray}
 \sum_{n=0}^{\infty}C_{2n}^{(-1/2)}(\omega)\lambda^{2n}=\frac 12(r_++r_-) \nonumber\\
-\sum_{n=0}^{\infty}C_{2n+1}^{(-1/2)}(\omega)\lambda^{2n+1}=\frac 12(r_+-r_-) \ ,
\end{eqnarray}
and therefore, by taking $\lambda=M/R$, the coordinates $\{r,y\}$ from (\ref{trans}) for this case are given by
\begin{eqnarray}
r&=&M+\frac R2 (r_++r_-) = M(x+1) \nonumber\\
y&=&\frac{R}{2M} (r_+-r_-) =y_p \ ,
\end{eqnarray}
where $\{x,y_p\}$ are the prolate spheroidal coordinates \cite{prolate}, \cite{algomio}.

This radial coordinate $r$ is easily recognized since it is merely the so-called {\it standard radial coordinate of Schwarzschild} and, as is known,  the $g_{00}$ component of the Schwarzschild metric written in this coordinates is $g_{00}=-1+2 M/r$, and hence the prescribed form of the metric component (\ref{u}) is recovered by this coordinate system and the relativistic Monopole Solution is described by  a function $u$ equal to the spherical Newtonian potential $M/r$.

In the following section we shall see the reasons why we refer to this system of coordinates as a Multipole-Symmetry Adapted one.

\section{Interpretation and characterization of MSA coordinates}

The function $u$ written in a MSA system of coordinates $\{r,y\}$ should be a solution of  equation (\ref{ernymsa}), (that is, the meaning of the condition (EI)), but at the same time it is also a solution of the following  system of differential equations
\begin{eqnarray}
0&=&r^2u_{rr}+2ru_r+(1-y^2)u_{yy}-2y u_{y}\equiv \bigtriangleup u\nonumber\\
0&=&\partial^{N+1}_y u \ ,
\label{nsym}
\end{eqnarray}
 for any value of $N$, whenever the function $u$ represents the Multipole Solution with a finite number
($N+1$) of RMM (\ref{u}):
\begin{equation}
 u_{MSA}=\sum_{n=0}^N\frac{M_n}{r^{n+1}}P_n(y)\ .
\label{umsa}
\end{equation}

Therefore, the symmetries of the system of differential equations (\ref{nsym}), obtained in NG \cite{cqg}, that allow one to extract from all solutions of the axially symmetric Laplace equation those with the prescribed Newtonian Multipole Moments work identically in the case of using MSA coordinates, but now the quantities $M_n$ are the RMM (\ref{umsa}). In this sense, the same family of  vector fields that  are the infinitesimal generators of certain one-parameter groups of transformations can be constructed. These vector fields represent symmetries of the  system of differential equations (\ref{nsym}) whose
group-invariant solutions   turn out  to be the family of axisymmetric potentials related to specific gravitational multipoles (\ref{umsa}).

The question we want to answer now is whether those groups of symmetry exist for the system of differential equations joined by (\ref{nsym}) and (\ref{ernymsa}); if  so, we could generalize the above-mentioned results to GR, establishing  a relationship between the  existence of a certain symmetry and the uniqueness of the solutions of the Einstein equations with a prescribed  multipole structure. Let us remark that  equation (\ref{ernymsa}) is the corresponding Ernst equation for the function $u$, and hence we could say more appropriately that Einstein equations admit the symmetry.

 Moreover, we wish to know whether the action of these symmetries  on the equation (\ref{ernymsa})
might provide conditions on the unknown functions $f_n(\omega)$ and $g_n(\omega)$ (\ref{trans}) to characterize the MSA coordinates.

\subsection{Multipole symmetries in GR}

Let
\begin{equation}
\textbf{v}= r\frac{\partial}{\partial r}-u\frac{\partial}{\partial u}
\label{uve}
\end{equation}
be a vector field  on an open subset $M \subset X \times U$, where $X={\mathbb {R}}^2$ is the space representing the independent variables,  coordinates $\{x\}=(r,y)$ being MSA coordinates such that (\ref{umsa}) is fulfilled for $N=0$,  and $U={\mathbb {R}}$,  with the coordinate $u$ that  represents the dependent variable.

We can state the following theorem:

\noindent {\bf Theorem 1}

The system of equations $\bigtriangleup_{\nu}(x,u^{(n)})=0$ given by
\begin{equation}
\left\{
\begin{array}{c}
\bigtriangleup_1(x,u^{(n)})  \equiv  \bigtriangleup u=0 \nonumber\\
\bigtriangleup_2(x,u^{(n)})  \equiv  u_{y}=0\nonumber\\
\bigtriangleup_3(x,u^{(n)}) \qquad  ,
\end{array}
\right.
\label{system1}
\end{equation}
where $\bigtriangleup_1$ is the Laplace equation (with axial symmetry)(\ref{nsym}), $\bigtriangleup_2$ the so-called supplementary equation \cite{cqg}, and $\bigtriangleup_3$   equation (\ref{ernymsa}), admits a symmetry group whose infinitesimal generator is {\bf v}.

\vspace*{5mm}

\noindent Proof:

The prolongation of {\bf v} acting on the supplementary equation is $pr^{(1)} \textbf{v}\left[\bigtriangleup_2\right]=- u_{y}$, and the second prolongation of this vector acting on $\bigtriangleup_1$ is $pr^{(2)} \textbf{v}\left[\bigtriangleup_1\right]= -\bigtriangleup u$. Therefore, both prolongations are zero whenever the system of equations  $\bigtriangleup_{\nu}(x,u^{(n)})=0$ is fulfilled.

With respect to the third equation of the system, it is straightforward to see that
\begin{eqnarray}
pr^{(2)} \textbf{v}\left[\bigtriangleup_3\right]&=&Du_r+Eu_y+r\left(A_ru_{rr}+B_ru_{yy}+2C_ru_{ry}+D_ru_r+E_ru_y\right)+\nonumber\\
& &-\frac{1}{2u-1}\left(Au_{rr}+Bu_{yy}+2Cu_{ry}+
Du_r+Eu_y\right)+\nonumber\\
& & -\frac{2r}{2u-1}\left(A_ru_r^2+B_ru_y^2+2C_ru_ru_y\right)-4\bigtriangleup_3\ .
\label{pro3}
\end{eqnarray}

Since the prolongations of the vector only need to vanish on solutions of the system of equations  (\ref{system1}) \cite{olver}, we make use of  equations $\bigtriangleup_2$ and $\bigtriangleup_3$ to obtain the following relation between the derivatives of $u$
\begin{equation}
u_{rr}=\frac{2}{2u-1}u_r^2-\frac DAu_r \ ,
\label{Eprevi}
\end{equation}
and we  substitute this and $\bigtriangleup_2$ into (\ref{pro3}) to obtain
\begin{equation}
pr^{(2)} \textbf{v}\left[\bigtriangleup_3\right]=u_r\left[rD_r-\frac DA (rA_r-A)\right]+u_r^2\left[-\frac{2A}{(2u-1)^2}\right]\ .
\label{18.2}
\end{equation}
The derivatives $u_r$ and $u_r^2$ are related by equations $\bigtriangleup_3$ and $\bigtriangleup_1$ (with $u_y=0$) as follows
\begin{equation}
u_r\left(-\frac{2A}{r}\right)=u_r^2\frac{2A}{2u-1}\ ,
\label{18.3}
\end{equation}
and therefore the second prolongation of $\bf v$ acting on $\bigtriangleup_3$ is
\begin{equation}
pr^{(2)} \textbf{v}\left[\bigtriangleup_3\right]=u_r\left[rD_r-\frac DA(rA_r-A)-\frac{1}{2u-1}(-\frac{2A}{r}+D)\right]\ .
\label{18.4}
\end{equation}
Equation (\ref{Eprevi}) is the Ernst equation for the function $u$ with the constraint $\bigtriangleup_2$. It should therefore be fulfilled by the Monopole Solution, represented by $u=M/r$, since we have required the system $\{r,y\}$ to be MSA coordinates for the Monopole case (spherical symmetry), and hence we can state  the following relation between the coefficients $A$ and $D$
\begin{equation}
\frac DA = \frac 2r \frac{r-M}{r-2M} \ .
\label{E}
\end{equation}
Finally, by using  equation (\ref{E}) and its derivative with respect to $r$, i.e.,
\begin{equation}
rD_r=r\frac DA\left(A_r-\frac Ar\right)+2A-2A\frac{r-M}{(r-2M)^2}
\end{equation}
in (\ref{18.4}), we have that:
\begin{equation}
pr^{(2)} \textbf{v}\left[\bigtriangleup_3\right]=-\frac{2AM}{r-2M}u_r\left[\frac{1}{r(2u-1)}+\frac{1}{r-2M}\right] \ .
\label{18.6}
\end{equation}
This expression is zero if at least one of the following conditions hold:
\begin{equation}
A=0 \ , \quad u_r=0 \ , \quad u=M/r \ .
\end{equation}
The first condition ($A=0$) can  obviously be neglected because it implies that  equation $\bigtriangleup_3$ disappears; the second condition means that $u=a+bf(y)$, $a$, and $b$ being arbitrary constants and $f(y)$ an arbitrary function of the variable $y$. Nevertheless, that expression is not a solution of the system (\ref{system1}), except for the case $u=a$, because $u_y \neq 0$, and, since  $u=M/r$ has been forced to be a solution of $\bigtriangleup_3$, which is a non-linear equation, then the linear combination $u=a+M/r$ is no longer a solution of $\bigtriangleup_3$.

Hence, we must finally conclude that  $pr^{(2)} \textbf{v}\left[\bigtriangleup_3\right]=0$ iff $u=M/r$, i.e., whenever
$\bigtriangleup_{\nu}(x,u^{(n)})=0$, as is the case, since that is the only solution of  system (\ref{system1}).

$\hfill{\square}$

\vspace*{1cm}
Let $\{r,y\}$ be an MSA system of coordinates such that (\ref{umsa}) is fulfilled for $N=1$. We can then state the following:

\noindent {\bf Theorem 2}

The system of equations $\bigtriangleup_{\nu}(x,u^{(n)})=0$ given by
\begin{equation}
\left\{
\begin{array}{c}
\bigtriangleup_1(x,u^{(n)})  \equiv  \bigtriangleup u=0 \\
\bigtriangleup_2(x,u^{(n)})  \equiv  u_{yy}=0\\
\bigtriangleup_3(x,u^{(n)}) \ ,
\end{array}
\right.
\label{system2}
\end{equation}
where $\bigtriangleup_1$ is the Laplace equation (with axial symmetry)(\ref{nsym}), $\bigtriangleup_2$ the so-called supplementary equation \cite{cqg}, and $\bigtriangleup_3$   equation (\ref{ernymsa}), admits a symmetry group whose infinitesimal generator is
\begin{equation}
 \textbf{v}= r\frac{\partial}{\partial r}+y\frac{\partial}{\partial y}-u\frac{\partial}{\partial u}\ .
\label{uve2}
\end{equation}

\vspace*{5mm}

\noindent Proof:

In \cite{cqg} the null conditions on the prolongations of the vector field (\ref{uve2}) acting on the first two equations of  system (\ref{system2}), $pr^{(2)} \textbf{v}\left[\bigtriangleup_1\right]=pr^{(2)} \textbf{v}\left[\bigtriangleup_2\right]=0$, whenever these two equations are fulfilled, were satisfied.
Now, we  explore whether $pr^{(2)} \textbf{v}\left[\bigtriangleup_3\right]$ vanishes for the solutions of the system (\ref{system2}).
It is straightforward to calculate  (see \cite{cqg} for details) that the second prolongation of vector (\ref{uve2}) acting on equation $\bigtriangleup_3$ is
\begin{eqnarray}
 pr^{(2)}
\textbf{v}\left[\bigtriangleup_3\right]&=&u_{rr}\left(rA_r+yA_y-A\right)+u_{yy}\left(rB_r+yB_y-B\right)+\nonumber\\
& &u_{ry}\left(2rC_r+2yC_y-2C\right)+\nonumber\\
& &u_r\left(rD_r+yD_y\right)+u_y\left(rE_r+yE_y\right)+\nonumber\\
& &u_ru_y\left[-\frac{4}{2u-1}\left(rC_r+yC_y\right)+2\frac{2u-2}{(2u-1)^2}2C\right]+\nonumber\\
& &u_r^2\left[-\frac{2}{2u-1}\left(rA_r+yA_y\right)+2\frac{2u-2}{(2u-1)^2}A\right]+\nonumber\\
& &u_y^2\left[-\frac{2}{2u-1}\left(rB_r+yB_y\right)+2\frac{2u-2}{(2u-1)^2}B\right]\ .
\label{19.1}
\end{eqnarray}

Since $pr^{(2)} \textbf{v}\left[\bigtriangleup_3\right]=0$  only needs to hold for the solutions of (\ref{system2}), we can substitute the derivative $u_{rr}$ from   equation $\bigtriangleup_1$,
\begin{equation}
u_{rr}=\frac{2}{A(2u-1)}\left(Au_r^2+Bu_y^2+2Cu_ru_y\right)-\frac{2C}{A}u_{ry}-\frac DAu_r-\frac EA u_y
\label{urr}
\end{equation}
into (\ref{19.1}) as follows
\begin{eqnarray}
 &&pr^{(2)}
\textbf{v}\left[\bigtriangleup_3\right]=u_{yy}\left(rB_r+yB_y-B\right)+\nonumber\\
& &\hskip 2cm u_{ry}\left[2rC_r+2yC_y-2C\left(r\frac{A_r}{A}+y\frac{A_y}{A}\right)\right]+\nonumber\\
& &\hskip 2cm u_r\left[rD_r+yD_y+D\left(1-r\frac{A_r}{A}+y\frac{A_y}{A}\right)\right]+\nonumber\\
& &u_y\left[rE_r+yE_y+E\left(1-r\frac{A_r}{A}+y\frac{A_y}{A}\right)\right]+u_r^2\left[-\frac{2A}{(2u-1)^2}\right]\nonumber\\
& &u_ru_y\left[-\frac{4}{2u-1}\left(rC_r+yC_y\right)+2\frac{2u-2}{(2u-1)^2}2C+\frac{4C}{2u-1}\left(r\frac{A_r}{A}+y\frac{A_y}{A}\right)\right]+\nonumber\\
& &u_y^2\left[-\frac{2}{2u-1}\left(rB_r+yB_y\right)+2\frac{2u-2}{(2u-1)^2}B+\frac{2B}{2u-1}\left(r\frac{A_r}{A}+y\frac{A_y}{A}\right)\right]\ .
\label{19.2}
\end{eqnarray}
By using  expression (\ref{urr}) in equations $\bigtriangleup_1$ and $\bigtriangleup_2$, the derivative $u_r^2$ should satisfy the following equation
\begin{equation}
 \frac{2}{2u-1}u_r^2=-\frac{2 \left[B u_y^2+2C u_ru_y\right]}{(2u-1)A}+2\frac CAu_{ry}-\left(\frac
2r-\frac DA\right)u_r+\left(\frac EA+\frac{2y}{r^2}\right)u_y\ .
\label{ur2}
\end{equation}
By replacing expressions (\ref{ur2}), (\ref{urr}) and  equation $\bigtriangleup_2$ in (\ref{19.2}) we have that
\begin{eqnarray}
 &&pr^{(2)}
\textbf{v}\left[\bigtriangleup_3\right]=\nonumber\\
&&u_{ry}\left[2rC_r+2yC_y-2C\left(r\frac{A_r}{A}+y\frac{A_y}{A}\right)-\frac{2C}{2u-1}\right]+\nonumber\\
&&u_r\left[rD_r+yD_y+D\left(1-r\frac{A_r}{A}+y\frac{A_y}{A}\right)+\frac{2A}{r(2u-1)}-\frac{D}{2u-1}\right]+\nonumber\\
&&u_y\left[rE_r+yE_y+E\left(1-r\frac{A_r}{A}+y\frac{A_y}{A}\right)+\frac{2A y}{r^2(2u-1)}-\frac{E}{2u-1}\right]+\nonumber\\
&&u_ru_y\left[-\frac{4}{2u-1}\left(rC_r+yC_y\right)+\frac{4C}{2u-1}\left(r\frac{A_r}{A}+y\frac{A_y}{A}\right)\right]+\nonumber\\
&&u_y^2\left[-\frac{2}{2u-1}\left(rB_r+yB_y\right)+\frac{2B}{2u-1}\left(r\frac{A_r}{A}+y\frac{A_y}{A}\right)\right]\ .
\label{19.3}
\end{eqnarray}

Since the coefficients of the various monomials in the first-order and second-order partial derivatives of $u$ in this expression must be equal to zero, we impose the following conditions on the coefficients and their derivatives of equation $\bigtriangleup_3$
\begin{eqnarray}
 C=0 \label{condis1}\\
A\left(rB_r+yB_y\right)=B\left(rA_r+yA_y\right)\\
rD_r+yD_y-\frac DA\left(rA_r+yA_y\right)+\frac{2u-2}{2u-1}D=-\frac{2A}{r(2u-1)}\\
rE_r+yE_y-\frac EA\left(rA_r+yA_y\right)+\frac{2u-2}{2u-1}E=\frac{2A y}{r^2(2u-1)}\ .
\label{condis}
\end{eqnarray}

The general solution of the system of equations $\bigtriangleup_1$ and $\bigtriangleup_2$ is as follows 
\begin{equation}
 u= c_1\frac{y}{r^2}+c_2 r y +c_3+\frac {c_4}{r}\ .
\label{general}
\end{equation}
Since $\{r,y\}$ is a MSA system of coordinates (for the multipole order considered) we can
 force equation $\bigtriangleup_3$ to possess the Monopole-Dipole Solution (among those from (\ref{general})
) represented by ${\bar u}=\frac 1rM+\frac{y}{r^2} M_1$ ($M$ and $M_1$ being the Monopole and Dipole moments respectively) and so, in addition to (\ref{condis1})-(\ref{condis}), we have the following condition ($C=0$)
\begin{equation}
 A\left[\bar u_{rr}-\frac{2}{2\bar u-1}\bar u_r^2\right]-\frac{2}{2\bar u-1}B \bar u_y^2+D\bar u_r+E\bar
u_y=0\ .
\label{otra}
\end{equation}
Moreover, this imposed condition implies that the only solution of the system of equations (\ref{system2}) is the Monopole-Dipole $\bar u$, and henceforth we can replace $\bar u$  by $u$ in equations (\ref{condis1})-(\ref{condis}), leading to the following expressions:
\begin{eqnarray}
&(a)& r\nu_r+y\nu_y=0 \nonumber\\
&(b)& r\kappa_r+y\kappa_y+\frac{2\bar u-2}{2\bar u-1} \kappa=-\frac{2}{r(2\bar u-1)} \nonumber\\
&(c)& r\mu_r+y\mu_y+\frac{2\bar u-2}{2\bar u-1} \mu=\frac{2y}{r^2(2\bar u-1)} \ ,
\label{condis2}
\end{eqnarray}
where $\nu\equiv\frac BA$, $\kappa\equiv\frac DA$ and $\mu\equiv\frac EA$. Equation (\ref{condis2}.a) requires  $\nu$ to be an arbitrary function of $y/r$ and the solutions of equations (\ref{condis2}.b), (\ref{condis2}.c) for $\kappa$, $\mu$ are the following functions
\begin{equation}
\nu=\nu(y/r) \ , \qquad \kappa=-\frac{2r+F_1(y/r)}{r^2(2\bar u-1)} \ , \qquad \mu=\frac{2y-F_2(y/r)}{r^2(2\bar u-1)}\ .
\label{kamu}
\end{equation}
Equivalently, these solutions of the determining equations (\ref{condis2}) can be written as follows:
\begin{eqnarray}
\nu A &=& B\nonumber\\
(2r+F_1)A &=&(r^2-2Mr-2M_1y)D\nonumber\\
(2y-F_2)A &=& -((r^2-2Mr-2M_1y)E\ .
\label{ABDE}
\end{eqnarray}
These functions $\nu$, $\kappa$ and $\mu$ are related by means of condition (\ref{otra}), leading to the following relation between $F_1$, $F_2$ and $\nu$
\begin{equation}
M_1 F_2=2M^2+4M_1^2\left(\frac yr\right)^2+8MM_1\frac yr-2M_1^2\nu+MF_1+2M_1F_1\frac yr\ .
\label{rela}
\end{equation}
The first equation of (\ref{ABDE}) and equation (\ref{rela}) allow us to write the following expression:
\begin{equation}
(2y-F_2)A=2M_1B+A\left[2y-\frac{2M^2}{M_1}-8M\frac yr-4M_1\left(\frac yr\right)^2-F_1\left(\frac{M}{M_1}+2\frac yr\right)\right]\ .
\label{esto}
\end{equation}
Finally, by substituting  expression (\ref{esto}), and the function $F_1$ obtained from the second equation of (\ref{ABDE}) in the third equation of (\ref{ABDE}), we have that:
\begin{eqnarray}
&&A\left[6y-2 \frac{M^2}{M_1}-8M\frac yr-4M_1\left(\frac yr\right)^2+2r\frac{M}{M_1}\right]+B 2 M_1\nonumber+\hskip 1cm\\
&&-D\left[(r^2-2Mr-2M_1y)\left(\frac{M}{M_1}+2\frac yr\right)\right]+E\left[(r^2-2Mr-2M_1y)\right]=0\ .\nonumber\\
\label{porfin}
\end{eqnarray}

We must therefore say that the second prolongation of the vector ${\bf v}$ acting on $\bigtriangleup_3$ vanishes whenever the system of equations (\ref{system2}) is fulfilled, iff  condition (\ref{porfin}) holds, and hence the proof of this theorem can be concluded since this condition(\ref{porfin}) is equivalent\footnote{Note that equation (\ref{otra}) is exactly equal to  expression (\ref{porfin}).} to saying that the Monopole-Dipole function $\bar u$ is  a solution of $\bigtriangleup_3$; that is, the assumption from the beginning  of the theorem if  coordinates used $\{r,y\}$ are MSA coordinates.

$\hfill{\square}$

\vspace*{1cm}
The relevance of the theorem comes from the relationship that can be established between the existence of the symmetry and the gauge of coordinates that provides the Newtonian form of the Monopole-Dipole Solution.

If we recall  definitions (\ref{lb1}-\ref{lb2}), then  the last expression (\ref{porfin}), in addition to  the first  equation  (\ref{condis})($C=0$), provides the following explicit conditions in the coordinate transformation:
\begin{eqnarray}
&(a)& \hat{LB}_1(r,y)\equiv R^2r_Ry_R+(1-\omega^2)r_{\omega}y_{\omega}=0\nonumber\\
&(b)& 2\hat{LB}_1(r,r)\left[-Mr^2(r-M)+M_1y(4Mr-3r^2+2yM_1)\right]=\nonumber\\
&&r(r^2-2Mr-2M_1y)\left[M_1r\hat{LB}_2(y)-(Mr+2yM_1)\hat{LB}_2(r)\right]+\nonumber\\
&&2M_1^2r^2\hat{LB}_1(y,y)\ .
\label{ejl}
\end{eqnarray}
Note that if we take $M_1=0$ in the above expression,  we obtain condition (\ref{E}), which must be used for the determination of the coordinates in the Monopole case. We shall discuss  these results in the following section.

\subsection{Characterization of the MSA systems of coordinates}
\vspace*{5mm}
{\bf A)The Monopole Solution}

From theorem 1, one can conclude that there exists a symmetry of the system of differential equations (\ref{system1}) iff we force  equation (\ref{Eprevi}) to posses a solution of the Monopole type $u=M/r$;  in other words, equation (\ref{E}) must be satisfied. Equation (\ref{Eprevi}) is the Ernst equation for the function $u$ with the constraints given by the other equations of the system (\ref{system1}), and it should be taken into account that the system of MSA coordinates that  we are using  allows us to characterize the relativistic Monopole Solution with a function $u$ written as the Newtonian Monopole. Therefore, since the only solution of  system (\ref{system1}) is the Monopole Solution, we can state that this system of differential equations admits a  symmetry group that can be related  to the uniqueness of the solution of the system by means of the existence of MSA coordinates.

Furthermore,  condition (\ref{E}) allows us to determine the gauge of coordinates in which the relativistic solution having only the Monopole Moment is given by a function with the same analytic form as the classical Monopole potential in NG. We proceed to do this in the following way. First, we substitute the coordinate transformation (\ref{trans}) in  condition (\ref{E}), which becomes a  constraint over the coordinate transformation, taking into account  definitions (\ref{lb1}-\ref{lb2}); i.e.
\begin{equation}
 r(r-2M) \hat{LB}_2(r)=2(r-M)\hat{LB}_1(r,r)\ .
\label{new}
\end{equation}
 Second, we solve the corresponding differential equations that appear at each order in the power series
expansion, and the uniqueness of the solution  is provided by the following boundary conditions:
\begin{equation}
 r(\omega=\pm1)=R+M \ ,
\label{bound}
\end{equation}
which implies that all functions $f_n(\omega)$ vanish for all $n > 1$ along the axis orthogonal to the equatorial plane.

As already noted, that radial coordinate $r$  is merely the Schwarzschild standard coordinate, the coordinate $y$ being free of constraints because of the spherical symmetry. The system of coordinates characterized by solving  equation (\ref{new}) with boundary conditions (\ref{bound}), can aptly be said to be adapted to the Monopole symmetry group, whose infinitesimal generator is (\ref{uve}), for several reasons. First, we  see that, written in these coordinates, the solution does not depend on the angular coordinate. However, in addition another feature contributes to characterizing these coordinates:
we refer to it as MSA because of the interrelation between the existence of the symmetry and the system of coordinates itself. Second, the function $u$ that describes the relativistic solution with a finite number of RMM acquires the form of the classical Multipole potential, and hence all the conclusions obtained for the Newtonian case can be assumed again for this function, which can be considered as the group-invariant solution of a system of differential equations (\ref{nsym}) that admits the symmetry.
\vspace*{5mm}

{\bf B)The Monopole-Dipole symmetry}

In analogy with the previous case, Theorem 2  allows us to establish a relationship between the existence of a symmetry of a certain system of equations and the MSA coordinates for the Monopole-Dipole Solution.
We have seen that the  uniqueness of the solution of the system of differential equations (\ref{system2}) can be deduced if  the Ernst equation for the function $u$ written in MSA coordinates is required to have a solution with the analytic form of the classical Monopole-Dipole gravitational potential. At the same time, this condition leads to the existence of a symmetry group for that system of equations.

The main goal obtained in the previous case (Monopole) is the  calculation of the coordinate $r$ by means of the constraint (\ref{new}) that Theorem 1 introduces in the  transformation of the coordinates; with appropriate boundary conditions (\ref{bound}), the choice of the functions $f_n(\omega)$ is unique and $r$ is fully determined.

Nevertheless, we cannot perform the complete determination of the MSA coordinates for the Monopole-Dipole case by using the constraints (\ref{ejl}) introduced by Theorem 2. Equation (\ref{ejl}.a) means that the new coordinates $r$ and $y$ preserve the orthogonality since they must be asymptotically Cartesian coordinates. This equation also implies that there are not cross terms in the metric written in MSA coordinates ($g_{ij}=0$, for $i\neq j$). 
If we substitute the prescribed gauge transformation  from Weyl coordinates (\ref{trans}) in constraint (\ref{ejl}.a), the corresponding series expansion leads to the following equations
\begin{equation}
 (1-\omega^2)\sum_{i=1}^k f^{\prime}_i(\omega)g^{\prime}_{k-i}(\omega) = k
g_k(\omega)\sum_{n=2}^{k-1}(n-1)(k-n)f_n(\omega)g_{k-n}(\omega)\ ,
\label{ecuages}
\end{equation}
where $g_0(\omega)=0$, the symbol $(^{\prime})$ denotes the derivative with respect to the variable $\omega$,  and $k$ is the different order of the expansion in the parameter $1/R$. This expression allows us  to write any function $g_k(\omega)$ in terms of the functions $f_n(\omega)$ and $g_n(\omega)$ of lower order ($n<k$) and their derivatives. As can be seen, the good behaviour of the functions $g_n(\omega)$ is recovered, since they must be zero along the axis of symmetry ($\omega=\pm1$).
With the expressions  of $g_k(\omega)$ obtained from (\ref{ecuages}), we may solve the corresponding equations at each order  of the series expansion of the other constraint, (\ref{ejl}.b), for the functions $f_n(\omega)$ alone. But now we do not have a suitable boundary condition to obtain a unique solution of the functions $f_n(\omega)$ for the complete determination of the coordinate $r$ for this Monopole-Dipole case.  We can demand that the limit $M_1=0$ must lead to the same functions as the Monopole case, but this  only allows us to determine the arbitrary constants of integration for $f_1(\omega)$ and $f_2(\omega)$, because from the next order onwards the equations involve  arbitrary constants, since the functions $f_n(\omega)$ for odd $n$ are null in the Monopole case.

With this procedure we obtain a family of coordinates that transforms the Ernst equation into another one that admits the function $\bar u$, representing the Monopole-Dipole Solution, as a solution. Nevertheless, the coordinate transformation is not unique.

\section{Conclusion}

The exterior gravitational field of an isolated and static compact body with axial symmetry is described in GR by means of the Weyl family of solutions, which depends on a set of arbitrary coefficients $\{a_n\}$ whose values  univocally determine each specific solution. If one is looking for solutions that are well known and physically meaningful,   it is necessary to relate this set of coefficients to the RMM in order to make a suitable selection of them.
If we work with  an MSA system of coordinates, then the solution that describes the gravitational behaviour of compact bodies with a prescribed multipole structure can be constructed by identifying the function $u$  with the Newtonian potential  and considering the constants of the classical potential  to be exactly the RMM of the solution. The transformation of a solution (Weyl) into another one (MSA) fixes the change of coordinates by requiring the function $u$ to be a solution of the corresponding Ernst equation.

Nevertheless, this procedure  needs, a priori, to know the set of coefficients $\{a_n\}$ of the desired Multipole Solution, although  the existence and explicit knowledge of these systems of coordinates is relevant enough and they become very useful, at least for the following topics: Application of this work could  shed light on  study of the influence and relevance of  different RMM in the behaviour of test particles along geodesics \cite{herrera} for different sources, since the MSA coordinates provide us with the exact Multipole Solutions in a very simple way. The deviation of the source from the spherical configuration is a very important feature, for example, for describing the fate of the collapse of self-gravitating systems  \cite{herrera2}. The calculation of circular geodesics at successive distances from the source can be used to determine its multipole structure. Additionally, some authors have attempted to  \cite{chusedu}  relate the RMM to  the structure of the source by means of quantities defined over the distribution tensor of the source. The existence of MSA coordinates for any Multipole Solution seems to be a very useful tool for achieving these aims in the frame of global stationary axisymmetric solutions of the
Einstein  equations.

Except for the Monopole case, in which the gauge is well-known (Schwarzs\-child radial coordinate), the MSA system of coordinates for the other cases are given by means of two series expansions in the inverse radial Weyl coordinate. However, if we work at large distances from the source, the approximate character of the coordinates  is negligible and the change of coordinates is completely determined. In addition, the function $u$ is an exact solution with the finite number of desired RMM. Also, we have defined a family of static and axially symmetric exact vacuum   solutions with a prescribed multipole structure in a system of coordinates defined with a suitable order of approximation.

Finally, this work affords another conclusion: two theorems have been proved that allow us to establish a relationship between the existence of a certain symmetry of  a system of differential equations (the correponding Ernst equation for the function $u$ is included among them) and the existence of a system of coordinates in which that function $u$ can be written  analytically equal to the Newtonian potential but in terms of the RMM. 

This result is relevant in itself, and some implications can be derived from it.
In particular,  the construction of the MSA system of coordinates for the Monopole case is supported by the proof of Theorem 1, without knowledge of the corresponding set of coefficients $\{a_n\}$, since the existence of the function $u$, or the MSA system of coordinates, is equivalent to the satisfaction of the corresponding Ernst equation by that function. This condition requires that the functions involved in the change of coordinates must satisfy  some differential equations whose solution is unique for suitable boundary conditions.  Accordingly, the existence of a one-parameter group of transformations can be stated, whose infinitesimal generator is (\ref{uve}), which represents a symmetry of the system of equations joined by the Laplace equation with axial symmetry, the supplementary equation and the Ernst equation for the function $u$ written in a system of coordinates adapted to that symmetry (MSA).

For the Monopole-Dipole case, Theorem 2 allow us to establish the same relationship between the existence of the Monopole-Dipole symmetry and the corresponding MSA coordinate system, although unfortunately the characterization of the gauge by means of the conditions provided by the theorem leads to a family of undefined coordinates in terms of arbitrary constants.

\section{Appendix A}

The following expressions show the first ten RMM of any Weyl solution in terms of its coefficients $a_n$:
\begin{eqnarray}
M_0&=&-a_0 \ , \ M_1=-a_1 \ , \ M_2=-a_2+\frac 13 a_0^3 \ , \ M_3=-a_3+a_1a_0^2\nonumber\\
M_4&=&\frac 87 a_2a_0^2-\frac{19}{105}a_0^5+\frac 67 a_0a_1^2-a_4 \nonumber \\
M_5&=&\frac 43 a_3a_0^2-\frac{19}{21}a_1a_0^4+\frac{12}{7}a_2a_0a_1-a_5+\frac 27 a_1^3\nonumber\\
M_6&=&\frac{20}{11}a_3a_0a_1-\frac{34}{21}a_0^3a_1^2-\frac{23}{21}a_2a_0^4+\frac{389}{3465}a_0^7+\frac{17}{11}a_4a_0^2+\frac{60}{77}a_0a_2^2+\nonumber\\
&+&\frac 67a_2a_1^2-a_6\nonumber\\
M_7&=&-\frac{206}{143}a_1^3a_0^2+\frac{389}{495}a_1a_0^6+\frac{120}{143}a_2^2a_1-\frac{595}{429}a_3a_0^4+\frac{126}{143}a_3a_1^2+\frac{23}{13}a_5a_0^2+\nonumber\\
&+&\frac{282}{143}a_4a_0a_1-a_7+\frac{20}{13}a_3a_0a_2-\frac{1504}{429}a_2a_1a_0^3\nonumber\\
M_{8}&=&\frac{2948}{1365}a_1^2a_0^5-\frac{652}{1001}a_0a_1^4+\frac{40}{143}a_2^3-\frac{257}{3465}a_0^9+\frac{226}{143}a_4a_0a_2+\nonumber\\
&-&\frac{4464}{1001}a_2a_1^2a_0^2-\frac{1744}{429}a_3a_1a_0^3-\frac{5204}{3003}a_2^2a_0^3+\frac{28}{39}a_0a_3^2-\frac{58}{33}a_4a_0^4+\nonumber\\
&+&\frac{12}{13}a_4a_1^2+2a_6a_0^2+\frac{44312}{45045}a_2a_0^6+\frac{240}{143}a_3a_2a_1+\frac{28}{13}a_5a_0a_1-a_8\nonumber\\
M_9&=& \frac{163508}{51051}a_0^4a_1^3-\frac{257}{385}a_1a_0^8+\frac{3192}{2431}a_3a_0^6+\frac{1988}{2431}a_3^2a_1+\frac{120}{143}a_3a_2^2+\nonumber\\
&-&\frac{486}{221}a_5a_0^4+\frac{216}{221}a_5a_1^2-\frac{11724}{2431}a_4a_1a_0^3+\frac{40}{17}a_6a_0a_1-\frac{44424}{17017}a_2a_0a_1^3+\nonumber\\
&-&\frac{10908}{85085}a_1^5+\frac{3430}{2431}a_4a_0a_3-\frac{11944}{2431}a_3a_1^2a_0^2-\frac{9080}{2431}a_3a_2a_0^3+\frac{366}{221}a_5a_0a_2+\nonumber\\
&+&\frac{418144}{85085}a_2a_1a_0^5+\frac{38}{17}a_7a_0^2+\frac{378}{221}a_4a_2a_1-a_9-\frac{5748}{1309}a_2^2a_1a_0^2\nonumber\\
M_{10}&=&-a_{10}+\frac{828}{323}a_7a_0a_1+\frac{10041124}{969969}a_2a_1^2a_0^4-\frac{10908}{17017}a_2a_1^4-\frac{17389}{20349}a_2a_0^8+\nonumber\\
&-&\frac{555820}{223839}a_1^2a_0^7+\frac{226580}{88179}a_2^2a_0^5-\frac{459700}{323323}a_2^3a_0^2-\frac{262556}{138567}a_3^2a_0^3+\nonumber\\
&+&\frac{3500}{4199}a_3^2a_2+\frac{12902}{7293}a_4a_0^6+\frac{39150}{46189}a_4a_2^2+\frac{30870}{46189}a_0a_4^2-\frac{2624}{969}a_6a_0^4+\nonumber\\
&+&\frac{336}{323}a_6a_1^2+\frac{47}{19}a_8a_0^2-\frac{193130}{46189}a_4a_2a_0^3+\frac{75180}{46189}a_4a_3a_1-\frac{896}{323}a_3a_0a_1^3+\nonumber\\
&-&\frac{94632}{24871}a_2^2a_0a_1^2+\frac{70412}{24871}a_1^4a_0^3+\frac{566}{323}a_6a_0a_2-\frac{257460}{46189}a_4a_1^2a_0^2+\nonumber\\
&+&\frac{5992}{4199}a_5a_0a_3-\frac{427568}{46189}a_3a_2a_1a_0^2+\frac{44152}{7293}a_3a_1a_0^5+\frac{7416}{4199}a_5a_2a_1+\nonumber\\
&+&\frac{443699}{8729721}a_0^{11}-\frac{24116}{4199}a_5a_1a_0^3
\label{A1}
\end{eqnarray}

From the above expressions we can extract, at each order, the corresponding coefficient $a_n$ in terms of the RMM:

\begin{eqnarray}
a_0&=&-M_0 \ , \ a_1=-M_1 \ , \ a_2=-\frac 13M_0^3-M_2 \ , \ a_3=-M_1M_0^2-M_3\nonumber\\
a_4&=&-\frac 87M_0^2M_2-\frac 15M_0^5-\frac 67 M_0M_1^2-M_4\nonumber\\
a_5&=&-\frac 43M_0^2M_3-\frac{12}{7}M_0M_1M_2-M_1M_0^4-\frac 27M_1^3-M_5\nonumber\\
a_6&=&-\frac{20}{11}M_1M_0M_3-\frac{25}{21}M_0^4M_2-\frac{38}{21}M_0^3M_1^2-\frac{17}{11}M_0^2M_4-\frac{60}{77}M_0M_2^2+\nonumber\\
&-&\frac{6}{7}M_1^2M_2-\frac{1}{7}M_0^7-M_6\nonumber\\
a_7&=&-M_1M_0^6-\frac{128}{33}M_0^3M_1M_2-\frac{49}{33}M_0^4M_3-\frac{120}{143}M_2^2M_1-\frac{18}{11}M_1^3M_0^2+\nonumber\\
&-&\frac{23}{13}M_0^2M_5-\frac{126}{143}M_1^2M_3-\frac{282}{143}M_0M_1M_4-M_7-\frac{20}{13}M_0M_3M_2\nonumber\\
a_8&=&-2M_0^2M_6-\frac{28}{39}M_3^2M_0-\frac{820}{429}M_2^2M_0^3-\frac{92}{33}M_1^2M_0^5-\frac{108}{143}M_0M_1^4+\nonumber\\
&-&\frac{1}{9}M_0^9-\frac{40}{33}M_2M_0^6-\frac{720}{143}M_2M_0^2M_1^2-\frac{1904}{429}M_3M_0^3M_1-\frac{266}{143}M_0^4M_4+\nonumber\\
&-&M_8-\frac{40}{143}M_2^3-\frac{226}{143}M_0M_4M_2-\frac{28}{13}M_0M_1M_5-\frac{240}{143}M_2M_1M_3+\nonumber\\
&-&\frac{12}{13}M_1^2M_4\nonumber\\
a_9&=&-\frac{120}{143}M_2^2M_3-M_9-\frac{366}{221}M_0M_5M_2-\frac{108}{715}M_1^5-\frac{784}{143}M_1^2M_0^2M_3+\nonumber\\
&-&\frac{432}{143}M_0M_1^3M_2-\frac{888}{143}M_0^5M_1M_2-\frac{224}{143}M_0^6M_3-\frac{30}{13}M_0^4M_5+\nonumber\\
&-&\frac{216}{221}M_1^2M_5-\frac{744}{143}M_0^3M_1M_4-\frac{378}{221}M_2M_1M_4-\frac{584}{143}M_0^3M_3M_2+\nonumber\\
&-&\frac{38}{17}M_0^2M_7-\frac{40}{17}M_0M_1M_6-\frac{604}{143}M_0^4M_1^3-\frac{3430}{2431}M_0M_3M_4+\nonumber\\
&-&\frac{708}{143}M_0^2M_1M_2^2-\frac{1988}{2431}M_1M_3^2-M_1M_0^8\nonumber\\
a_{10}&=&-\frac{336}{323}M_1^2M_6-\frac{14940}{2431}M_1^2M_0^2M_4-\frac{11010}{2431}M_0^3M_4M_2+\nonumber\\
&-&\frac{1356}{221}M_0^3M_1M_5-\frac{75180}{46189}M_1M_4M_3-\frac{5992}{4199}M_0M_5M_3-\frac{48}{17}M_0^4M_6+\nonumber\\
&-&\frac{7416}{4199}M_1M_5M_2-\frac{294}{143}M_0^6M_4-\frac{5012}{2431}M_0^3M_3^2-\frac{47}{19}M_0^2M_8+\nonumber\\
&-&\frac{548}{143}M_1^4M_0^3-\frac{566}{323}M_0M_2M_6-\frac{828}{323}M_0M_1M_7-\frac{108}{143}M_1^4M_2+\nonumber\\
&-&\frac{10728}{2431}M_1^2M_0M_2^2-M_{10}-\frac{540}{143}M_1^2M_0^7-\frac{30870}{46189}M_0M_4^2+\nonumber\\
&-&\frac{460}{143}M_0^5M_2^2-\frac{176}{17}M_1M_0^2M_2M_3-\frac{39150}{46189}M_2^2M_4+\nonumber\\
&-&\frac{148}{11}M_1^2M_0^4M_2-\frac{1064}{143}M_1M_0^5M_3-\frac{7728}{2431}M_1^3M_0M_3+\nonumber\\
&-&\frac{175}{143}M_2M_0^8-\frac{300}{187}M_0^2M_2^3-\frac{1}{11}M_0^{11}-\frac{3500}{4199}M_2M_3^2
\label{A2}
\end{eqnarray}

\section{Appendix B}

The following expressions show the functions $f_n(\omega)$ appearing in (\ref{trans}), for the Monopole case. Since the spherical symmetry only requires to define the radial coordinate, and the function $u$ does not depend on $y$, the condition (E0)  provides itself this result:
\begin{eqnarray}
& &f_1(\omega)= M \ , \ f_2(\omega)= -\frac 12 M^2 (-1+\omega^2) \ , \ f_3(\omega)= 0 \nonumber\\
& &f_4(\omega)= -\frac 18 M^4 (5 \omega^4+1-6 \omega^2) \ , \ f_5(\omega)= 0 \nonumber\\
& &f_6(\omega)= -\frac{1}{16} M^6 (-35 \omega^4-1+15 \omega^2+21 \omega^6) \ , \ f_7(\omega)= 0 \nonumber \\
& & f_8(\omega)= -\frac{1}{128} M^8 (5+429 \omega^8-924 \omega^6+630 \omega^4-140 \omega^2) \ , \ f_9(\omega)=0 \nonumber\\
& & f_{10}(\omega)= \frac{1}{256} M^{10} (7-2431 \omega^{10}+6435 \omega^8-6006 \omega^6+2310 \omega^4-315 \omega^2)\nonumber\\
\label{B1}
\end{eqnarray}

For a more general case, we have calculated the MSA coordinates  for the solution having only the  Monopole, Dipole and Cuadrupole moments, and the functions $f_n(\omega)$ and $g_n(\omega)$, up to order $10$, are the following:

\begin{eqnarray}
f_1(\omega)&=&M_0 \ , \ 
f_2(\omega)=-\frac12 M_0^2 \omega^2+\frac12 M_0^2\nonumber\\
f_3(\omega)&=&M_0 M_1 \omega (1-\omega^2)+\frac12 M_2 (1-3\omega^2)+\frac{M_1^2}{M_0} \omega^2\nonumber\\
f_4(\omega)&=&-\frac{5}{42} M_1^2 \omega^2-\frac83 \frac{M_1^3}{M_0^2} \omega^3-\frac54 M_0 M_2 \omega^4+\frac{39}{14} \omega^2 M_0 M_2-3 M_1 \omega \frac{M_2}{M_0}+\nonumber\\
&+&\frac{1}{84} M_1^2+\frac18 M_0^4 (6\omega^2-1-5\omega^4)+7 M_1 \omega^3 \frac{M_2}{M_0}-\frac34 M_1^2 \omega^4+\nonumber\\
&-&\frac{19}{28} M_0 M_2+\frac23 \frac{M_1^3}{M_0^2} \omega\nonumber\\
f_5(\omega)&=&-\frac{7}{2}  \omega^2\frac{M_1^4}{M_0^3}+\frac{M_2^2}{M_0}+5 M_0^3 M_1 \omega^3+\frac{71}{21}  \omega^3\frac{M_1^3}{M_0}+2 M_0 M_1^2 \omega^4+\nonumber\\
&-&\frac{7}{2} M_0^3 M_1 \omega^5-9  \omega^2\frac{M_2^2}{M_0}+2 M_0^2 M_2 \omega^2-\frac{13}{21}  \omega\frac{M_1^3}{M_0}-\frac{3}{2} M_1 \omega M_0^3+\nonumber\\
&-&3 M_0^2 M_2 \omega^4+12  \omega^4\frac{M_2^2}{M_0}+\frac{22}{3}  \omega^4\frac{M_1^4}{M_0^3}-\frac{3}{4} \frac{M_1^2 M_2}{M_0^2}-M_1^2 \omega^2 M_0+\nonumber\\
&+&\frac{1}{6} \frac{M_1^4}{M_0^3}-\frac{103}{4} M_1^2 \omega^4 \frac{M_2}{M_0^2}-\frac{87}{14} M_1 M_2 \omega^3+\frac{31}{2} M_1^2 \omega^2 \frac{M_2}{M_0^2}+\nonumber\\
&+&\frac{85}{28} M_1 M_2 \omega-\frac{9}{4} M_1 M_2 \omega^5\nonumber\\
f_6(\omega)&=&-\frac{13}{105} \frac{M_1^4}{M_0^2}-\frac{3293}{840} M_1^2 \omega^2 M_0^2+\frac{1777}{168} M_0^2 M_1^2 \omega^4+\frac{223}{840} M_0^2 M_1^2+\nonumber\\
&-&\frac{21}{16} M_0^6 \omega^6+\frac{46}{5} \omega M_2 \frac{M_1^3}{M_0^3}-\frac{1247}{616} M_2^2+\frac{1}{16} M_0^6+\nonumber\\
&-&\frac{308}{15} \omega^5 \frac{M_1^5}{M_0^4}-\frac{1378}{105} \frac{M_1^4}{M_0^2}\omega^4-\frac{63}{8} M_0^2 M_1^2 \omega^6+\frac{214}{15} \frac{M_1^5 \omega^3}{M_0^4}+\nonumber\\
&-&\frac{26}{15} \frac{M_1^5 \omega}{M_0^4}-\frac{15}{8} M_2^2 \omega^6-\frac{13089}{616} M_2^2 \omega^4+\frac{10347}{616} M_2^2 \omega^2+\nonumber\\
&-&\frac{45}{8} M_0^3 M_2 \omega^6-4 M_1^3 \omega^5+\frac{496}{105} \frac{M_1^4 \omega^2}{M_0^2}+\frac{25}{28} \frac{M_1^2 M_2}{M_0}+\nonumber\\
&-&\frac{6}{5} M_1 \omega M_0 M_2+\frac{901}{10} \frac{M_2 M_1^3 \omega^5}{M_0^3}-\frac{713}{10} \frac{M_2 M_1^3 \omega^3}{M_0^3}+15 M_0 M_1 M_2 \omega^5+\nonumber\\
&+&\frac{6597}{140} \frac{M_1^2 \omega^4 M_2}{M_0}-\frac{247}{20} \frac{M_2^2 M_1 \omega}{M_0^2}+\frac{366}{5} \frac{M_2^2 M_1 \omega^3}{M_0^2}+\nonumber\\
&-&\frac{1617}{20} \frac{M_2^2 M_1 \omega^5}{M_0^2}-\frac{1781}{70} \frac{M_1^2 \omega^2 M_2}{M_0}+\frac{35}{16} \omega^4 M_0^6-\frac{2}{15} M_1^3 \omega^3+\nonumber\\
&+&\frac{2}{15} M_1^3 \omega-\frac{15}{16} \omega^2 M_0^6-\frac{361}{56} M_0^3 M_2 \omega^2+\frac{705}{56} M_0^3 M_2 \omega^4+\frac{73}{168} M_0^3 M_2\nonumber\\
&-&\frac{29}{5} M_1 \omega^3 M_0 M_2\nonumber\\
f_7(\omega)&=&M_1^2 \omega^2 M_0^3-\frac{142687}{13860} M_1^3 \omega^3 M_0+\frac{17}{2} M_0^4 M_2 \omega^4-2 M_0^4 M_2 \omega^2+\nonumber\\
&+&\frac{98}{9} \frac{M_1^6 \omega^2}{M_0^5}-\frac{475}{9} \frac{M_1^6 \omega^4}{M_0^5}-\frac{144}{5} \frac{M_2^3 \omega^2}{M_0^2}-\frac{8503}{315} \frac{M_1^5 \omega^3}{M_0^3}+\nonumber\\
&-&\frac{33}{4} M_0 M_1^3 \omega^7-\frac{99}{8} M_0^5 M_1 \omega^7+\frac{2618}{45} \frac{M_1^6 \omega^6}{M_0^5}-\frac{88049}{9240} M_1 \omega^3 M_0^2 M_2+\nonumber\\
&+&\frac{13747}{13860} M_1^3 \omega M_0-\frac{105}{8} M_0^5 M_1 \omega^3-\frac{15}{2} M_0^4 M_2 \omega^6+5 M_0^3 M_1^2 \omega^6+\nonumber\\
&-&\frac{839}{630} \frac{M_1^4 \omega^2}{M_0}+\frac{49}{3} \frac{M_1^4 \omega^6}{M_0}+\frac{107}{60} \frac{M_1^4 M_2}{M_0^4}-\frac{191}{60} \frac{M_1^2 M_2^2}{M_0^3}+\nonumber\\
&+&\frac{1889}{84} M_0 M_1^3 \omega^5+\frac{107}{45} \frac{M_1^5 \omega}{M_0^3}-\frac{8783}{20} \frac{M_2^2 \omega^4 M_1^2}{M_0^3}+\frac{15329}{315} \frac{M_1^5 \omega^5}{M_0^3}+\nonumber\\
&-&\frac{414}{5} \frac{M_2^3 \omega^6}{M_0^2}-\frac{123}{140} M_1^2 M_2+\frac{73}{126} \frac{M_1^4 \omega^4}{M_0}+\frac{981}{10} \frac{M_2^3 \omega^4}{M_0^2}+\nonumber\\
&+&\frac{189}{8} M_0^5 M_1 \omega^5+\frac{81}{2} M_0 M_2^2 \omega^6+\frac{3}{2} \frac{M_2^3}{M_0^2}+\frac{2}{15} \frac{M_1^4}{M_0}+\nonumber\\
&-&\frac{13}{45} \frac{M_1^6}{M_0^5}+\frac{61}{28} M_0 M_2^2-5 M_1^2 \omega^4 M_0^3+\frac{1583}{105} M_1^2 \omega^2 M_2+\nonumber\\
&+&\frac{1305}{56} M_0^2 M_1 M_2 \omega^5-\frac{18353}{60} \frac{M_2 \omega^6 M_1^4}{M_0^4}+\frac{1627}{4} \frac{M_2^2 \omega^6 M_1^2}{M_0^3}+\nonumber\\
&-&\frac{3947}{60} \frac{M_2 M_1^4 \omega^2}{M_0^4}-\frac{897742}{5005} \frac{M_1 \omega^3 M_2^2}{M_0}+\frac{4147313}{20020} \frac{M_1 \omega^5 M_2^2}{M_0}+\nonumber\\
&+&\frac{16793}{572} \frac{M_1 \omega M_2^2}{M_0}-\frac{209}{8} M_0^2 M_1 M_2 \omega^7-\frac{121}{2} M_1^2 M_2 \omega^6+\nonumber\\
&+&\frac{1303}{12} \frac{M_2^2 M_1^2 \omega^2}{M_0^3}-\frac{7681}{35} \frac{M_2 M_1^3 \omega^5}{M_0^2}+\frac{2155}{14} \frac{M_2 M_1^3 \omega^3}{M_0^2}+\nonumber\\
&+&\frac{15}{8} M_0^5 M_1 \omega-\frac{2711}{140} M_0 M_2^2 \omega^4-\frac{38}{5} M_0 M_2^2 \omega^2-\frac{159}{10} \frac{M_2 M_1^3 \omega}{M_0^2}\nonumber\\
&+&\frac{23719}{9240} M_1 \omega M_0^2 M_2+\frac{2047}{420} M_1^2 \omega^4 M_2+\frac{18113}{60} \frac{M_2 M_1^4 \omega^4}{M_0^4}\nonumber\\
f_8(\omega)&=&\frac{35}{32} \omega^2 M_0^8-\frac{429}{128} M_0^8 \omega^8-\frac{315}{64} M_0^8 \omega^4+\frac{231}{32} M_0^8 \omega^6+\nonumber\\
&+&\frac{1318143}{112112} M_1^4 \omega^4-\frac{26323}{840} M_1^4 \omega^6-\frac{927191}{840840} M_1^4 \omega^2-\frac{383}{1056} M_2 M_0^5+\nonumber\\
&-&\frac{138851}{24024} \frac{M_2^3}{M_0}+\frac{127}{315} \frac{M_1^6}{M_0^4}-\frac{48658}{2205} \frac{M_1^6 \omega^2}{M_0^4}-\frac{227}{70} \frac{M_1^4 M_2}{M_0^3}+\nonumber\\
&-&\frac{936731}{672672} M_0^2 M_2^2-\frac{3655}{7392} M_0^4 M_1^2+\frac{19486}{105} \frac{M_1^7 \omega^5}{M_0}^6-\frac{1156}{21} \frac{M_1^7 \omega^3}{M_0^6}+\nonumber\\
&+&\frac{166}{45} \frac{M_1^7 \omega}{M_0^6}-\frac{117}{32} M_1^4 \omega^8-\frac{13553439}{40040} \frac{M_2^3 \omega^4}{M_0}+\frac{62329}{8008} \frac{M_2^2 M_1^2}{M_0^2}+\nonumber\\
&+&\frac{431143}{1681680} M_0 M_2 M_1^2-\frac{333}{35} M_0^3 \omega^3 M_1 M_2+27 M_0^3 \omega^7 M_1 M_2+\nonumber\\
&-&\frac{36}{5} M_0^3 \omega^5 M_1 M_2+\frac{12}{7} \omega M_0^3 M_1 M_2+\frac{196349}{840} M_1^2 \omega^6 M_0 M_2\nonumber\\
&+&\frac{52341}{520} \frac{M_2^3 \omega^2}{M_0}+\frac{892359}{3080} \frac{M_2^3 \omega^6}{M_0}+\frac{292162}{2205} \frac{M_1^6 \omega^4}{M_0^4}-56 \frac{M_1^5 \omega^7}{M_0^2}+\nonumber\\
&-&\frac{1496}{9} \frac{M_1^7 \omega^7}{M_0^6}-\frac{27494}{2205} \frac{M_1^5 \omega^5}{M_0^2}-\frac{3662}{2205} \frac{M_1^5 \omega}{M_0^2}-\frac{4897}{112} M_0^5 \omega^4 M_2+\nonumber\\
&-&\frac{30004069}{210210} M_1^2 \omega^4 M_0 M_2+\frac{50233}{3640} \omega^2 M_0 M_2 M_1^2-\frac{715}{32} M_0^5 \omega^8 M_2+\nonumber\\
&+&\frac{1595}{28} M_0^5 \omega^6 M_2+\frac{4825}{462} \omega^2 M_0^5 M_2+\frac{736}{105} M_0^2 \omega^3 M_1^3-\frac{188}{15} M_0^2 \omega^5 M_1^3+\nonumber\\
&-&\frac{35827}{560} M_0^4 \omega^4 M_1^2-\frac{1573}{32} M_0^4 \omega^8 M_1^2+\frac{8305}{84} M_0^4 \omega^6 M_1^2+\nonumber\\
&-&\frac{10}{21} \omega M_0^2 M_1^3+\frac{15916}{1155} \omega^2 M_0^4 M_1^2-\frac{85223}{1121120} M_1^4-\frac{5}{128} M_0^8+\nonumber\\
&-&\frac{385723}{2205} \frac{M_1^6 \omega^6}{M_0^4}+\frac{34096}{2205} \frac{M_1^5 \omega^3}{M_0^2}-\frac{715}{32} M_0^2 M_2^2 \omega^8+\nonumber\\
&-&\frac{503}{14} \frac{M_2^3 M_1 \omega}{M_0^3}-\frac{612686}{735} \frac{M_2 M_1^4 \omega^4}{M_0^3}+\frac{75697}{490} \frac{M_2 M_1^4 \omega^2}{M_0^3}+\nonumber\\
&-&\frac{25330}{21} \frac{M_2 M_1^5 \omega^5}{M_0^5}-\frac{7866797}{25480} \frac{M_2^2 M_1^2 \omega^2}{M_0^2}+\frac{10523}{735} \frac{M_1^3 \omega M_2}{M_0}+\nonumber\\
&+&\frac{17111}{735} \frac{M_1^3 \omega^5 M_2}{M_0}+\frac{148747}{980} M_1 \omega^3 M_2^2+\frac{79313}{35} \frac{M_2^2 M_1^3 \omega^5}{M_0^4}+\nonumber\\
&-&\frac{29103}{245} \frac{M_1^3 \omega^3 M_2}{M_0}+\frac{56453}{140} \frac{M_2^3 M_1 \omega^3}{M_0^3}-\frac{38004}{35} \frac{M_2^3 M_1 \omega^5}{M_0^3}+\nonumber\\
&+&\frac{6221}{245} M_1 \omega^5 M_2^2-\frac{1077}{4} M_1 \omega^7 M_2^2-\frac{663}{16} M_0 M_2 M_1^2 \omega^8+\nonumber\\
&-&\frac{22771}{30} \frac{M_2^2 M_1^3 \omega^3}{M_0^4}+\frac{34285679}{560560} \omega^4 M_0^2 M_2^2-\frac{1143841}{140140} \omega^2 M_0^2 M_2^2+\nonumber\\
&+&\frac{112353}{140} \frac{M_2^3 \omega^7 M_1}{M_0^3}+\frac{35619}{35} \frac{M_2 \omega^7 M_1^5}{M_0^5}-\frac{4121}{77} \omega^6 M_0^2 M_2^2+\nonumber\\
&+&\frac{5858}{105} \frac{\omega M_2^2 M_1^3}{M_0^4}-\frac{904}{35} \frac{\omega M_2 M_1^5}{M_0^5}+\frac{1123}{3} \frac{\omega^3 M_2 M_1^5}{M_0^5}+\nonumber\\
&+&\frac{374531797}{280280} \frac{M_2^2 M_1^2 \omega^4}{M_0^2}+265 \frac{M_1^3 \omega^7 M_2}{M_0}-\frac{73619837}{56056} \frac{M_2^2 M_1^2 \omega^6}{M_0^2}+\nonumber\\
&+&\frac{695879}{735} \frac{M_2 M_1^4 \omega^6}{M_0^3}-\frac{125499}{70} \frac{M_2^2 \omega^7 M_1^3}{M_0^4}-\frac{18163}{490} M_1 \omega M_2^2\nonumber\\
f_9(\omega)&=&\frac{3786191}{72072} M_1^3 \omega^3 M_0^3+\frac{105}{4} M_1 \omega^3 M_0^7-\frac{693}{8} M_1 \omega^5 M_0^7+\nonumber\\
&+&9 M_1^2 \omega^4 M_0^5+\frac{215}{2} M_1^2 \omega^4 M_0^2 M_2+\frac{3655803}{40040} M_1 M_0^4 M_2 \omega^3+\nonumber\\
&-&\frac{2139}{140} M_1^2 \omega^2 M_0^2 M_2-\frac{848053}{96096} M_1 M_0^4 M_2 \omega-\frac{1714497}{6160} M_1 M_0^4 M_2 \omega^5+\nonumber\\
&+&\frac{401}{20} M_2^2 M_0^3 \omega^2-\frac{1800157}{9240} M_0^3 M_1^3 \omega^5-\frac{354703}{90090} M_0^3 M_1^3 \omega+\nonumber\\
&+&2 M_2 M_0^6 \omega^2-16 M_2 M_0^6 \omega^4-\frac{35}{16} M_1 \omega M_0^7+\frac{1}{7} M_1^2 M_0^2 M_2+\nonumber\\
&+&\frac{253}{315} M_1^4 \omega^2 M_0-\frac{15772}{15} \frac{M_1^4 \omega^8 M_2}{M_0^2}-\frac{752747631}{224224} \frac{M_2^3 \omega^7 M_1}{M_0^2}+\nonumber\\
&+&\frac{14495}{56} M_0^3 M_1^3 \omega^7-\frac{10449}{140} \frac{M_2^4 \omega^2}{M_0^3}-\frac{22469}{1960} \frac{M_2^2 M_1^2}{M_0}+\nonumber\\
&-&\frac{1306727}{2205} \frac{M_1^7 \omega^5}{M_0^5}+\frac{13149}{28} \frac{M_2^4 \omega^4}{M_0^3}+\frac{4301}{9} \frac{M_1^8 \omega^8}{M_0^7}-M_1^2 \omega^2 M_0^5+\nonumber\\
&-&\frac{934429}{280} \frac{M_2 \omega^8 M_1^6}{M_0^6}-\frac{1271}{315} M_1^4 \omega^4 M_0+\frac{136}{11} M_2^3-\frac{2}{21} M_2^2 M_0^3+\nonumber\\
&+&\frac{231}{2} M_0^3 M_2^2 \omega^8+\frac{30763}{84084} \frac{M_1^5 \omega}{M_0}+34 M_0^6 M_2 \omega^6+\frac{119}{3} M_0 M_1^4 \omega^8+\nonumber\\
&+&+\frac{1807363}{2940} \frac{M_1^7 \omega^7}{M_0^5}-\frac{113867}{180} \frac{M_1^8 \omega^6}{M_0^7}+\frac{20725}{84} \frac{M_1^8 \omega^4}{M_0^7}+\nonumber\\
&+&\frac{608}{147} \frac{M_1^4 M_2}{M_0^2}+\frac{1621}{168} \frac{M_2^2 M_1^4}{M_0^5}-\frac{5189}{560} \frac{M_2^3 M_1^2}{M_0^4}-\frac{35887}{1260} \frac{M_1^8 \omega^2}{M_0^7}+\nonumber\\
&+&\frac{210779}{1260} \frac{M_1^5 \omega^7}{M_0}+\frac{50191}{2520} \frac{M_1^6 \omega^2}{M_0^3}-\frac{71303}{8820} \frac{M_1^7 \omega}{M_0^5}-\frac{769}{210} \frac{M_1^6 M_2}{M_0^6}+\nonumber\\
&+&\frac{20085}{56} M_0^4 M_1 M_2 \omega^7-\frac{5805579}{1120} \frac{M_2^3 \omega^8 M_1^2}{M_0^4}+\frac{662141}{4410} \frac{M_1^7 \omega^3}{M_0^5}+\nonumber\\
&+&\frac{819555}{112} \frac{M_2^2 \omega^8 M_1^4}{M_0^5}-\frac{578719153}{3363360} \frac{M_2^2 M_1^3 \omega}{M_0^3}-105 M_0^2 M_2 M_1^2 \omega^8+\nonumber\\
&-&\frac{6673937}{5880} \frac{M_2 M_1^5 \omega^3}{M_0^4}+\frac{8809390459}{3363360} \frac{M_2^2 M_1^3 \omega^3}{M_0^3}-\frac{405}{16} M_1^3 M_2 \omega^9+\nonumber\\
&-&\frac{9431951569}{1121120} \frac{M_2^2 M_1^3 \omega^5}{M_0^3}-\frac{12523453}{1681680} M_1^3 M_2 \omega+\frac{429}{4} M_0^7 M_1 \omega^7+\nonumber\\
&+&-\frac{715}{16} M_0^7 M_1 \omega^9-\frac{316}{315} M_0 M_1^4 \omega^6-\frac{1011517}{8820} \frac{M_1^6 \omega^4}{M_0^3}+\nonumber\\
&+&-\frac{7761}{20} M_2^3 \omega^8-\frac{845}{8} M_0^3 M_1^3 \omega^9+\frac{85089}{2695} M_2^3 \omega^6+\frac{4086767}{10780} M_2^3 \omega^4+\nonumber\\
&+&\frac{42251}{504} \frac{M_1^6 \omega^6}{M_0^3}+\frac{1148139}{140140} \frac{M_1^5 \omega^3}{M_0}-\frac{1125}{16} M_0 M_1 M_2^2 \omega^9+\nonumber\\
&-&\frac{18563603}{252252} \frac{M_1^5 \omega^5}{M_0}+\frac{2816}{15} \frac{M_1^6 \omega^8}{M_0^3}-\frac{132543}{140} \frac{M_2^4 \omega^6}{M_0^3}+\nonumber\\
&+&\frac{20493}{35} \frac{M_2^4 \omega^8}{M_0^3}-21 M_0^6 M_2 \omega^8-\frac{437691}{2695} M_2^3 \omega^2+14 M_0^5 M_1^2 \omega^8+\nonumber\\
&-&21 M_0^5 M_1^2 \omega^6+\frac{2546307}{5720} M_1^3 \omega^5 M_2+\frac{2536}{35035} M_1^3 \omega^3 M_2+\nonumber\\
&-&\frac{79123}{105} M_1^3 \omega^7 M_2-\frac{11393363}{2940} \frac{M_2 \omega^7 M_1^5}{M_0^4}+\frac{1259203}{1470} \frac{M_1^4 \omega^4 M_2}{M_0^2}+\nonumber\\
&+&\frac{990122241}{224224} \frac{M_2^3 M_1 \omega^5}{M_0^2}-\frac{33193}{420} M_0^2 M_1^2 \omega^6 M_2-\frac{259}{5} M_2^2 M_0^3 \omega^4+\nonumber\\
&-&\frac{5655}{32} M_0^4 M_1 M_2 \omega^9-\frac{35031}{98} \frac{M_2 \omega^6 M_1^4}{M_0^2}+\frac{69329927}{480480} \frac{M_2^3 M_1 \omega}{M_0^2}+\nonumber\\
&-&\frac{126526}{735} \frac{M_2 M_1^4 \omega^2}{M_0^2}+\frac{12186841}{2940} \frac{M_2 M_1^5 \omega^5}{M_0^4}-\frac{1353}{28} M_0^3 M_2^2 \omega^6+\nonumber\\
&-&\frac{164152063}{101920} \frac{M_2^3 M_1 \omega^3}{M_0^2}+\frac{2991113}{10780} \frac{M_2^2 M_1^2 \omega^6}{M_0}+\frac{3097}{2} \frac{M_2^2 M_1^2 \omega^8}{M_0}+\nonumber\\
&+&\frac{42701053}{1681680} M_0 M_1 \omega M_2^2-\frac{19308785}{12936} \frac{M_2^2 M_1^2 \omega^4}{M_0}+\nonumber\\
&-&\frac{951347}{1680} \frac{M_2^2 M_1^4 \omega^2}{M_0^5}+\frac{388861}{5880} \frac{M_2 M_1^5 \omega}{M_0^4}+\frac{186671}{840} \frac{M_1^6 M_2 \omega^2}{M_0^6}+\nonumber\\
&-&\frac{31}{315} M_0 M_1^4-\frac{341}{735} \frac{M_1^6}{M_0^3}+\frac{9}{4} \frac{M_2^4}{M_0^3}-\frac{314309}{168} \frac{M_1^6 M_2 \omega^4}{M_0^6}+\nonumber\\
&+&\frac{83}{180} \frac{M_1^8}{M_0^7}-\frac{4065127}{1120} \frac{M_2^3 \omega^4 M_1^2}{M_0^4}+\frac{40869853}{40040} M_0 M_1 \omega^7 M_2^2+\nonumber\\
&+&\frac{8890071}{1120} \frac{M_2^3 \omega^6 M_1^2}{M_0^4}-\frac{5957607}{560} \frac{M_2^2 \omega^6 M_1^4}{M_0^5}+\frac{653883}{8008} M_0 M_1 \omega^3 M_2^2+\nonumber\\
&+&\frac{557573}{1120} \frac{M_2^3 M_1^2 \omega^2}{M_0^4}+\frac{7630153}{1680} \frac{M_2^2 M_1^4 \omega^4}{M_0^5}-\frac{28687558}{35035} M_0 M_1 \omega^5 M_2^2+\nonumber\\
&+&\frac{3497168}{8085} \frac{M_2^2 M_1^2 \omega^2}{M_0}+\frac{722625097}{101920} \frac{M_2^2 \omega^7 M_1^3}{M_0^3}+\frac{1294039}{280} \frac{M_2 M_1^6 \omega^6}{M_0^6}\nonumber\\
f_{10}(\omega)&=&-\frac{65998679}{3811808} M_0 M_2^3-\frac{315}{256} M_0^{10} \omega^2-\frac{1522198}{2205} \frac{M_1^6 \omega^8}{M_0^2}\nonumber\\
&-&\frac{1091827871}{168168} \frac{M_2^3 M_1 \omega^5}{M_0}+\frac{116182867}{1261260} \frac{M_1^4 M_2 
\omega^2}{M_0}+\nonumber\\
&-&\frac{823063229}{140140} M_1^2 M_2^2 \omega^8-\frac{1071}{16} M_1^2 M_2^2 \omega^{10}-657 M_0^2 M_2^2 M_1 \omega^9+\nonumber\\
&+&\frac{28295301319}{6126120} M_1^2 M_2^2 \omega^6+\frac{86424015889}{3363360} \frac{M_1^2 \omega^8 M_2^3}{M_0^3}+\nonumber\\
&-&\frac{90647147471}{257297040} M_1^2 M_2^2 \omega^2+\frac{41574419}{2144142} M_1^2 M_2^2\omega^4+\nonumber\\
&-&\frac{510812297}{26460} \frac{M_1^6 \omega^6 M_2}{M_0^5}+\frac{720726269}{2522520} \frac{M_1^3 \omega M_2^2}{M_0^2}+\nonumber\\
&+&\frac{24253205017}{1441440} \frac{M_1^2 \omega^4 M_2^3}{M_0^3}-\frac{67749616547}{30270240} \frac{M_1^2\omega^2 M_2^3}{M_0^3}+\nonumber\\
&+&\frac{37401809}{5292} \frac{M_1^6 \omega^4 M_2}{M_0^5}-\frac{57147915281}{3027024} \frac{M_1^4 \omega^4 M_2^2}{M_0^4}+\nonumber\\
&-&\frac{8637393}{1120}\frac{M_1 \omega^5 M_2^4}{M_0^4}+\frac{44623451}{17640} \frac{M_1^5 \omega^7 M_2}{M_0^3}+\nonumber\\
&+&\frac{2124423}{160} \frac{M_1 \omega^7 M_2^4}{M_0^4}-\frac{165\
9717}{224} \frac{M_1 \omega^9 M_2^4}{M_0^4}+\nonumber\\
&+&\frac{16666453}{10584} \frac{M_1^5 \omega^3 M\
M_2}{M_0^3}+\frac{14932814329}{315315} \frac{M_1^4 \omega^6 M_2^2}{M_0^4}\nonumber\\
&+&\frac{1482330167}{687960} \frac{M_1^4 \omega^2 M_2^2}{M_0^4}
+\frac{201899419}{13230} \frac{M_1^6 \omega^8 M_2}{M_0^5}+\nonumber\\
&-&\frac{47883811}{8820} \frac{M_1^5 \omega^5 M_2}{M_0^3}-\frac{49533}{560} \frac{M_1 \omega M_2^4}{M_0^4}+\nonumber\\
&-&\frac{25556833901}{672672} \frac{M_1^2 \omega^6 M_2^3}{M_0^3}+\frac{240257}{60} \frac{M_1^5 \omega^9 M_2}{M_0^3}+\nonumber\\
&+&\frac{1801203}{1120} \frac{M_1 \omega^3 M_2^4}{M_0^4}+\frac{2995594751}{252252} \frac{M_1^3 \omega^5 M_2^2}{M_0^2}+\nonumber\\
&-&\frac{305701}{40} \frac{M_1^3 \omega^9 M_2^2}{M_0^2}-\frac{407164048}{105105} \frac{M_1^3 \omega^7 M_2^2}{M_0^2}+\nonumber\\
&-&\frac{18105}{32} M_0^3 M_1^2 M_2 \omega^{10}+\frac{23308919}{840} \frac{M_1^3 \omega^9 M_2^3}{M_0^5}+\nonumber\\
&-&\frac{11923937}{420} \frac{M_1^5 \omega^9 M_2^2}{M_0^6}+\frac{13639837}{1260} \frac{M_1^7 \omega^9 M_2}{M_0^7}+\nonumber\\
&-&\frac{2442373}{26460} \frac{M_1^5 \omega M_2}{M_0^3}-\frac{983324}{1323} \frac{M_1^6 \omega^2 M_2}{M_0^5}+\nonumber\\
&-&\frac{5623987}{3780} \frac{M_1^7 \omega^3 M_2}{M_0^7}+\frac{6494269}{756} \frac{M_1^7 \omega^5 M_2}{M_0^7}+\frac{15103}{1470} \frac{M_1^6 M_2}{M_0^5}+\nonumber\\
&+&\frac{33154333}{7560} \frac{M_1^5 \omega^3 M_2^2}{M_0^6}-\frac{209051}{1080} \frac{M_1^5 \omega M_2^2}{M_0^6}-\frac{29207791}{550368} M_1^2 \omega^2 M_0^3 M_2+\nonumber\\
&+&\frac{235639}{1008} \frac{M_1^3 \omega M_2^3}{M_0^5}+\frac{12025}{189} \frac{M_1^7 \omega M_2}{M_0^7}-\frac{8260429}{32340} \omega^3 M_0^2 M_2^2 M_1+\nonumber\\
&+&\frac{65076171}{160160} \frac{M_2^4 \omega^2}{M_0^2}+\frac{56525}{82368} M_0^6 M_1^2-\frac{2664951647}{1681680} M_1^2 \omega^6  M_0^3 M_2+\nonumber\\
&+&\frac{465748042}{945945} M_1^2 \omega^4 M_0^3 M_2+\frac{3401}{105} M_1 \omega^3 M_0^5 M_2+\frac{259}{5} M_1 \omega^7 M_0^5 M_2+\nonumber\\
&-&\frac{3296}{35} M_1 \omega^5 M_0^5 M_2+\frac{156327}{40} \frac{M_2^3 M_1 \omega^9}{M_0}-\frac{982936967}{30270240} \frac{M_2^2 M_1^4}{M_0^4}+\nonumber\\
&+&\frac{29351}{16} M_1^2 \omega^8 M_0^3 M_2-\frac{42882109}{72765} M_1^3 \omega^5 M_0 M_2+\nonumber\\
&+&\frac{4225037}{48510} M_1^3 \omega^3 M_0 M_2+\frac{604591}{1470} M_1^3 \omega^7  M_0 M_2+\nonumber\\
&-&\frac{50027}{9702} \omega M_0^2 M_2^2 M_1+\frac{24439994}{24255} \omega^5 M_0^2 M_2^2 M_1-\frac{21011}{72765} \omega M_0 M_2 M_1^3+\nonumber\\
&+&28 M_1 \omega^9 M_0^5 M_2-\frac{212}{105} M_0^5 M_1 M_2 \omega-\frac{479261}{980} \omega^7 M_0^2 M_2^2 M_1+\nonumber\\
&+&\frac{7}{256} M_0^{10}-\frac{81941597}{1513512} \frac{M_1^4 M_2 \omega^4}{M_0}+\frac{128708191}{105105} \frac{M_2^3 M_1 \omega^7}{M_0}+\nonumber\\
&+&\frac{60156797}{181621440} M_0^2 M_1^4-\frac{1452013}{72765} M_1^5 \omega^3-\frac{290663}{2205} M_1^5 \omega^7+\nonumber\\
&+&\frac{2346121}{24255} M_1^5 \omega^5+\frac{44089}{72765} M_1^5 \omega-\frac{546}{5} M_1^3 \omega^7 M_0^4+\nonumber\\
&+&\frac{8482}{105} M_1^3 \omega^5 M_0^4-\frac{5542}{315} M_1^3 \omega^3 M_0^4-\frac{677833}{1120} M_1^2 \omega^6 M_0^6+\nonumber\\
&+&\frac{106269}{448} M_1^4 \omega^8 M_0^2-\frac{2246545753}{181621440} M_1^4 \omega^2 M_0^2-\frac{799877851}{3363360} M_1^4 \omega^6 M_0^2+\nonumber\\
&+&\frac{1744579433}{18162144} M_1^4 \omega^4 M_0^2+\frac{8444941}{36960} M_1^2 \omega^4 M_0^6+\frac{25144057}{1681680} \omega^4 M_0^4 M_2^2+\nonumber\\
&+&\frac{893555}{1344} M_1^2 \omega^8 M_0^6+\frac{1422352}{35035} \omega^6 M_0^4 M_2^2+\frac{21271059}{12320} M_0 M_2^3 \omega^8+\nonumber\\
&+&\frac{12075043}{308880} \frac{M_2^3 M_1^2}{M_0^3}-\frac{1275}{8} M_0^4 M_2^2 \omega^{10}+\nonumber\\
&+&\frac{65870521}{26460} \frac{M_1^8 \omega^6}{M_0^6}-\frac{352}{3} M_1^5 \omega^9-\frac{11235935}{5292} \frac{M_1^8 \omega^8}{M_0^6}+\nonumber\\
&-&\frac{8764561}{19845} \frac{M_1^7 \omega^7}{M_0^4}-\frac{5873111}{1681680} \frac{M_1^4 M_2}{M_0}-\frac{27874}{45} \frac{M_1^7 \omega^9}{M_0^4}+\nonumber\\
&-&\frac{38337021}{12320} \frac{M_2^4 \omega^8}{M_0^2}+\frac{10355481}{2080} \frac{M_2^4 \omega^6}{M_0^2}-\frac{13368209}{840840} \omega^2 M_0^4 M_2^2+\nonumber\\
&-&\frac{17022571101}{9529520} M_0 M_2^3 \omega^6+\frac{780547}{7392} \omega^4 M_0^7 M_2+\nonumber\\
&+&\frac{214}{315} M_0^4 M_1^3 \omega+\frac{112}{3 M_1^3} \omega^9 M_0^4-\frac{57863}{224} \omega^6 M_0^7 M_2+\nonumber\\
&-&\frac{2866051}{192192} \omega^2 M_0^7 M_2+\frac{174555}{2464} \omega^8 M_0^4 M_2^2+\frac{13022941}{15135120} M_0^3 M_1^2 M_2+\nonumber\\
&-&\frac{88845511}{2882880} M_1^2 \omega^2 M_0^6+\frac{114075}{448} \omega^8 M_0^7 M_2+\frac{18619}{20384} M_0^4 M_2^2+\nonumber\\
&+&\frac{2935}{9152} M_2 M_0^7-\frac{8415}{64} M_0^2 M_1^4 \omega^{10}-\frac{5525}{64} M_0^7 M_2 \omega^{10}+\nonumber\\
&-&\frac{16575}{64} M_0^6 M_1^2 \omega^{10}+\frac{767417513}{64324260} M_1^2 M_2^2-\frac{78404639}{3783780} \frac{M_1^6 \omega^4}{M_0^2}+\nonumber\\
&+&\frac{6585608633}{57177120} M_0 M_2^3 \omega^2+\frac{929535058}{2837835} \frac{M_1^6 \omega^6}{M_0^2}+\nonumber\\
&-&\frac{1275}{32} M_0 M_2^3 \omega^{10}+\frac{18322867}{26460} \frac{M_1^7 \omega^5}{M_0^4}-\frac{189551}{34398 }\frac{M_1^6 \omega^2}{M_0^2}+\nonumber\\
&+&\frac{2690985929}{9529520} M_0 M_2^3 \omega^4-\frac{111826}{81} \frac{M_1^9 \omega^9}{M_0^8}-\frac{337931}{1890} \frac{M_1^7 \omega^3}{M_0^4}+\nonumber\\
&+&\frac{6788489}{79380} \frac{M_1^8 \omega^2}{M_0^6}-\frac{13638059}{15876} \frac{M_1^8 \omega^4}{M_0^6}+\nonumber\\
&+&\frac{768721}{79380} \frac{M_1^7 \omega}{M_0^4}+\frac{98062}{567} \frac{M_1^9 \omega^3}{M_0^8}-\frac{20558}{2835} \frac{M_1^9 \omega}{M_0^8}+\nonumber\\
&-&\frac{107608}{105} \frac{M_1^9 \omega^5}{M_0^8}+\frac{855242}{405} \frac{M_1^9 \omega^7}{M_0^8}-\frac{79368645}{32032} \frac{M_2^4 \omega^4}{M_0^2}+\nonumber\\
&-&\frac{2431}{256} M_0^{10} \omega^{10}-\frac{22003}{19845} \frac{M_1^8}{M_0^6}+\frac{6435}{256} M_0^{10} \omega^8+\nonumber\\
&+&\frac{1155}{128} M_0^{10} \omega^4-\frac{3003}{128}  M_0^{10} \omega^6+\frac{4107659}{11351340} \frac{M_1^6}{M_0^2}+\nonumber\\
&-&\frac{218931}{16016} \frac{M_2^4}{M_0^2}+\frac{131451\
323}{5040} \frac{M_1^3 \omega^5 M_2^3}{M_0^5}-\frac{15999047}{336} \frac{M_1^3 \omega^7 M_2^3}{M_0^5}+\nonumber\\
&+&\frac{39279559}{840} \frac{M_1^5 \omega^7 M_2^2}{M_0^6}-\frac{25118767}{5040} \frac{M_1^3 \omega^3 M_2^3}{M_0^5}+\nonumber\\
&-&\frac{184244621}{7560} \frac{M_1^5 \omega^5 M_2^2}{M_0^6}-\frac{1343113868}{315315} \frac{M_1^3 \omega^3 M_2^2}{M_0^2}+\nonumber\\
&-&\frac{6356431271}{183456 }\frac{M_1^4 \omega^8 M_2^2}{M_0^4}-\frac{21620803}{1260} \frac{M_1^7 \omega^7 M_2}{M_0^7}+\nonumber\\
&+&\frac{83616411}{28028} \frac{M_2^3 M_1 \omega^3}{M_0}-\frac{25940077}{84084} \frac{M_2^3 M_1\omega}{M_0}+\nonumber\\
&+&662 M_0 M_2 M_1^3 \omega^9+\frac{46612879}{11760} \frac{M_1^4 M_2 \omega^8}{M_0}-\frac{705291019}{291060} \frac{M_1^4 M_2 \omega^6}{M_0}\nonumber\\
\label{B7}
\end{eqnarray}

\begin{eqnarray}
g_1(\omega)&=&0 \ , \
g_2(\omega)=-(1-\omega^2)\frac 12 \omega M_0^2\nonumber\\
g_3(\omega)&=&-(1-\omega^2) \frac{1}{3 M_0}(3M_0^2 M_1 \omega^2-2 M_1^2 \omega+3\omega M_0 M_2- M_0^2 M_1)\nonumber\\
g_4(\omega)&=&-(1-\omega^2)\frac{1}{168 M_0^2}(126 M_1^2 \omega^3 M_0^2+147 \omega^3 M_0^6+210 M_0^3 M_2 \omega^3+\nonumber\\
&-&882 M_1 \omega^2 M_0 M_2+
336 M_1^3 \omega^2-234 M_0^3 M_2 \omega+10 M_0^2 M_1^2 \omega+\nonumber\\
&-&63 \omega M_0^6+126 M_1 M_0 M_2-28 M_1^3)\nonumber\\
g_5(\omega)&=&-(1-\omega^2)\frac{1}{420 M_0^3}(945 M_1 \omega^4 M_0^3 M_2+1890 M_0^6 M_1 \omega^4+52 M_0^2 M_1^3+\nonumber\\
&+&1260 M_0^5 M_2 \omega^3-840 M_1^2 \omega^3 M_0^4-2464 M_1^4 \omega^3-4032 M_0^2 M_2^2 \omega^3+\nonumber\\
&+&8652 M_2 M_1^2 M_0 \omega^3+1566 M_1 \omega^2 M_0^3 M_2-852 M_0^2 M_1^3 \omega^2+\nonumber\\
&-&1400 M_0^6 M_1 \omega^2+1512 M_0^2 M_2^2 \omega-336 M_0^5 M_2 \omega+588 M_1^4 \omega+\nonumber\\
&-&2604 M_2 M_1^2 M_0 \omega+224 M_1^2 \omega M_0^4+126 M_1 M_0^6-255 M_1 M_0^3 M_2)\nonumber\\
g_6(\omega)&=&-(1-\omega^2)\frac{1}{55440 M_0^4}(16016 M_1^5+948640 M_1^5 \omega^4-395472 M_1^5 \omega^2+\nonumber\\
&+&197736 M_1 \omega^2 M_0^5 M_2-831600 M_1 \omega^4 M_0^5 M_2+114114 M_2^2 M_1 M_0^2+\nonumber\\
&+&485056 M_1^4 \omega^3 M_0^2-99616 M_1^4 \omega M_0^2-85008 M_1^3 M_2 M_0+\nonumber\\
&+&103950 \omega^5 M_2^2 M_0^4+221760 M_1^3 \omega^4 M_0^4+6776 M_1^3 \omega^2 M_0^4+\nonumber\\
&-&512820 M_0^7 M_2 \omega^3-310410 \omega M_2^2 M_0^4+785340 \omega^3 M_2^2 M_0^4+\nonumber\\
&+&113190 M_0^7 M_2 \omega+381150 M_0^7 M_2 \omega^5+533610 M_0^6 M_1^2 \omega^5+\nonumber\\
&+&78386 M_0^6 M_1^2 \omega+11088 M_1 M_0^5 M_2-448140 M_1^2 \omega^3 M_0^6+\nonumber\\
&-&103950 \omega^3 M_0^{10}+17325 \omega M_0^{10}+114345 \omega^5 M_0^{10}-7392 M_1^3 M_0^4+\nonumber\\
&+&488664 M_1^2 \omega M_2 M_0^3-1741608 M_1^2 \omega^3 M_2 M_0^3+\nonumber\\
&+&1976436 M_1^3 M_2 \omega^2 M_0-2029104 M_2^2 M_1 \omega^2 M_0^2+\nonumber\\
&-&4162620 M_1^3 M_2 \omega^4 M_0+3735270 M_2^2 M_1 \omega^4 M_0^2)\nonumber\\
g_7(\omega)&=&-(1-\omega^2)\frac{1}{2522520 M_0^5}(-125789664 M_1^6 \omega^5+76076000 M_1^6 \omega^3+\nonumber\\
&-&7847840 M_1^6 \omega-176138820 M_1^3 M_2 \omega^2 M_0^3+2044328 M_1^4 \omega^3 M_0^4+\nonumber\\
&+&77882805 M_0^7 M_1 \omega^6 M_2-16109808 M_1^2 \omega^3 M_2 M_0^5-348842 M_0^6 M_1^3+\nonumber\\
&-&47750703 M_0^7 M_1 \omega^4 M_2-12415260 M_1^2 \omega M_2 M_0^5+\nonumber\\
&+&152612460 M_1^2 \omega^5 M_2 M_0^5+10947807 M_1 \omega^2 M_0^7 M_2+\nonumber\\
&-&41201160 M_1^4 \omega^5 M_0^4+1770912 M_1^4 \omega M_0^4+6450444 M_1^3 M_2 M_0^3+\nonumber\\
&-&10579590 M_2^2 M_1 M_0^4+24594570 M_0^6 M_1^3 \omega^6-15639624 M_0^9 M_2 \omega^3\nonumber\\
&-&87681880 M_1^5 \omega^4 M_0^2+32785896 M_1^5 \omega^2 M_0^2+179026848 M_2^3 \omega^5 M_0^3+\nonumber\\
&-&141405264 M_2^3 \omega^3 M_0^3+20756736 M_2^3 \omega M_0^3-1017016 M_1^5 M_0^2+\nonumber\\
&-&45939894 M_0^6 M_1^3 \omega^4+4926636 M_0^6 \omega M_2^2+22702680 M_0^9 M_2 \omega^5+\nonumber\\
&+&45090045 M_0^{10} M_1 \omega^6-54339285 M_0^{10} M_1 \omega^4+15786771 M_0^{10} M_1 \omega^2+\nonumber\\
&-&15135120 M_0^8 M_1^2 \omega^5-102162060 M_0^6 \omega^5 M_2^2-693381 M_1 M_0^7 M_2+\nonumber\\
&+&36159552 M_0^6 \omega^3 M_2^2-1153152 M_1^2 \omega M_0^8+12803830 M_1^3 \omega^2 M_0^6+\nonumber\\
&+&1729728 M_0^9 M_2 \omega+10426416 M_1^2 \omega^3 M_0^8-373258170 M_2^2 M_1 \omega^4 M_0^4+\nonumber\\
&+&47411364 M_1^4 M_2 \omega M_0+633008376 M_2^2 M_1^2 \omega^3 M_0^2+\nonumber\\
&-&78258180 M_2^2 M_1^2 \omega M_0^2-435146712 M_2 M_1^4 \omega^3 M_0+\nonumber\\
&+&395417880 M_1^3 M_2 \omega^4 M_0^3+661368708 M_2 M_1^4 \omega^5 M_0+\nonumber\\
&-&879458580 M_2^2 M_1^2 \omega^5 M_0^2+197155512 M_2^2 M_1 \omega^2 M_0^4-675675 M_0^{10} M_1)\nonumber\\
g_8(\omega)&=&-(1-\omega^2)\frac{1}{40360320 M_0^6}(1499806308 M_1^2 \omega^3 M_0^{10}+\nonumber\\
&+&170270100 M_1^4 \omega^7 M_0^6-2115133020 M_0^{11} M_2 \omega^5+\nonumber\\
&+&398670272 M_1^3 \omega^4 M_0^8-149112236 M_1^2 \omega M_0^{10}+\nonumber\\
&-&1343785212 M_0^8 M_2^2 \omega^3-170218048 M_0^8 M_1^3 \omega^2+89716788 M_0^8 M_2^2 \omega+\nonumber\\
&+&14477068 M_1^4 \omega M_0^6+175258512 M_2^2 M_1 M_0^6-341059716 M_1^4 \omega^3 M_0^6+\nonumber\\
&-&101474100 M_0^{11} M_2 \omega+1763421660 M_0^8 M_2^2 \omega^5-12684672 M_1 M_0^9 M_2+\nonumber\\
&+&1047642596 M_1^4 \omega^5 M_0^6+2705402700 M_0^{10} M_1^2 \omega^7+\nonumber\\
&-&3787563780 M_0^{10} M_1^2 \omega^5+121396275 M_0^{14} \omega^3+\nonumber\\
&+&1229728500 M_0^{11} M_2 \omega^7+964863900 M_0^{11} M_2 \omega^3+14240512 M_1^5 M_0^4+\nonumber\\
&+&833152320 M_1^7 \omega^2+5870184320 M_1^7 \omega^6+1040539500 M_0^8 \omega^7 M_2^2\nonumber\\
&-&4681316640 M_1^7 \omega^4-18610592 M_1^7-315630315 M_0^{14} \omega^5+\nonumber\\
&+&8456448 M_0^8 M_1^3+225450225 M_0^{14} \omega^7-11036025 M_0^{14} \omega+\nonumber\\
&+&1929727800 M_0^7 M_1^2 \omega^7 M_2+3321323688 M_2 M_1^2 M_0^7 \omega^3+\nonumber\\
&-&189787224 M_2 M_1^2 M_0^7 \omega-7883403528 M_2 M_1^2 M_0^7 \omega^5+\nonumber\\
&-&1271350080 M_1 \omega^6 M_0^9 M_2+103975872 M_1 \omega^2 M_0^9 M_2+\nonumber\\
&+&400912512 M_1 \omega^4 M_0^9 M_2-1821197664 M_2^2 M_1 \omega^4 M_0^6+\nonumber\\
&+&10867016160 M_2^2 M_1 \omega^6 M_0^6-2186245776 M_2^2 M_1 \omega^2 M_0^6+\nonumber\\
&+&261449760 M_1^6 \omega M_0^2-2945401888 M_1^6 \omega^3 M_0^2+130306176 M_1^5 M_2 M_0+\nonumber\\
&-&281465184 M_1^3 M_2^2 M_0^2+63608688 M_1^5 \omega^4 M_0^4+5295205344 M_1^6 \omega^5 M_0^2+\nonumber\\
&+&181261080 M_2^3 M_1 M_0^3+2260177920 M_1^5 \omega^6 M_0^4-316277104 M_1^5 \omega^2 M_0^4+\nonumber\\
&-&96782400 M_1^3 M_2 M_0^5+6885419688 M_2^3 \omega^3 M_0^5-1009570716 M_2^3 \omega M_0^5+\nonumber\\
&-&5665579920 M_1^5 M_2 \omega^2 M_0+30426396000 M_2 M_1^5 \omega^4 M_0+\nonumber\\
&+&576521088 M_1^3 M_2 \omega^4 M_0^5+63314747496 M_2^2 \omega^6 M_1^3 M_0^2+\nonumber\\
&-&35939855952 M_2 M_1^5 \omega^6 M_0-1751809488 M_1^4 M_2 \omega M_0^3+\nonumber\\
&-&8770104252 M_2^3 \omega^5 M_0^5+27390242880 M_2^3 \omega^4 M_1 M_0^3+\nonumber\\
&+&11488060584 M_2^2 M_1^3 \omega^2 M_0^2+3265593804 M_2^2 M_1^2 \omega M_0^4+\nonumber\\
&-&28341268956 M_2^3 \omega^6 M_1 M_0^3-6103020924 M_2^3 M_1 \omega^2 M_0^3+\nonumber\\
&-&57162465360 M_2^2 M_1^3 \omega^4 M_0^2-28659080736 M_2 M_1^4 \omega^5 M_0^3+\nonumber\\
&+&17840332224 M_2 M_1^4 \omega^3 M_0^3+1916785728 M_1^3 M_2 \omega^2 M_0^5+\nonumber\\
&-&27760378680 M_2^2 M_1^2 \omega^3 M_0^4+39754711980 M_2^2 M_1^2 \omega^5 M_0^4+\nonumber\\
&-&10695484800 M_1^3 M_2 \omega^6 M_0^5)\nonumber\\
g_9(\omega)&=&-(1-\omega^2)\frac{1}{90810720 M_0^7}(-15275704 M_1^5 M_0^6+36900864 M_0^{12} M_1^2 \omega+\nonumber\\
&+&5450156712 M_1 \omega^4 M_0^14-1821910868 M_1^3 \omega^2 M_0^{10}+\nonumber\\
&+&2542700160 M_0^{13} M_2 \omega^7+35239099896 M_2^3 \omega^7 M_0^7\nonumber\\
&+&12543230700 M_0^{10} M_1^3 \omega^8+1347869952 M_2^2 M_0^{10} \omega^3+\nonumber\\
&+&98163208 M_1^7 M_0^2-4116752640 M_0^8 M_1^4 \omega^7+\nonumber\\
&-&876431556 M_1 \omega^2 M_0^{14}+105656346 M_1^3 M_2 M_0^7+\nonumber\\
&-&327742272 M_2^2 M_0^{10} \omega-55351296 M_2 M_0^{13} \omega+\nonumber\\
&-&38575496960 M_1^8 \omega^7-2924105184 M_0^{13} M_2 \omega^5+\nonumber\\
&+&574766192 M_1^8 \omega+5449384512 M_0^{10} M_2^2 \omega^5+\nonumber\\
&+&964240992 M_0^8 M_1^4 \omega^5-22746463740 M_0^{10} M_1^3 \omega^6+\nonumber\\
&+&874377504 M_2 M_0^{13} \omega^3+38297571312 M_1^8 \omega^5+\nonumber\\
&-&4484584104 \omega M_2 M_1^6 M_0+81237765 M_1 M_0^{11} M_2+\nonumber\\
&-&1695133440 M_0^{12} M_1^2 \omega^7-10575665100 M_0^{14} M_1 \omega^6+\nonumber\\
&+&2237664 M_0^8 M_1^4 \omega+3009066840 M_2^3 \omega M_0^7+\nonumber\\
&+&6271615350 M_0^{14} M_1 \omega^8-166664346 M_0^8 M_1 M_2^2+\nonumber\\
&-&11987015040 M_0^{10} M_2^2 \omega^7+11804986908 M_0^{10} M_1^3 \omega^4+\nonumber\\
&+&71001216 M_1^4 \omega^3 M_0^8-7268072760 M_2^3 \omega^5 M_0^7+\nonumber\\
&-&13562878680 M_2^3 \omega^3 M_0^7+1949403456 M_0^{12} M_1^2 \omega^5+\nonumber\\
&-&9957948000 M_1^8 \omega^3-582918336 M_1^2 \omega^3 M_0^{12}+\nonumber\\
&+&95484445056 M_2 \omega^7 M_1^4 M_0^5+22072050 M_1 M_0^{14}+\nonumber\\
&+&3621679776 M_0^9 M_2 M_1^2 \omega^5+10897286400 M_0^9 M_2 M_1^2 \omega^7+\nonumber\\
&+&20985789825 M_0^{11} M_1 \omega^8 M_2-31206455280 M_0^{11} M_1 \omega^6 M_2+\nonumber\\
&+&48082385880 M_0^8 M_1 M_2^2 \omega^4-79406203500 M_0^8 M_1 M_2^2 \omega^6+\nonumber\\
&+&7236479250 M_0^8 M_1 M_2^2 \omega^8-29285259180 M_1^3 M_2 \omega^4 M_0^7+\nonumber\\
&+&992103912 M_1^3 M_2 \omega^2 M_0^7+2605132530 M_1^3 M_2 \omega^8 M_0^7+\nonumber\\
&+&58534088184 M_1^3 M_2 \omega^6 M_0^7-3834874116 M_0^8 M_1 M_2^2 \omega^2+\nonumber\\
&-&4918591392 M_1^2 \omega^3 M_0^9 M_2-2773705080 M_1 M_0^{11} M_2 \omega^2+\nonumber\\
&+&532975872 M_1^2 \omega M_0^9 M_2+15794246754 M_1 M_0^{11} M_2 \omega^4+\nonumber\\
&-&772894980 M_1^5 M_2 M_0^3-47263087872 \omega^7 M_2^4 M_0^4+\nonumber\\
&-&5233976176 M_1^7 \omega^2 M_0^2-579659080 M_1^6 \omega M_0^4+\nonumber\\
&-&13021279128 M_1^5 \omega^6 M_0^6-2494612120 M_1^6 \omega^5 M_0^4+\nonumber\\
&-&315702232 M_1^5 \omega^2 M_0^6-43420088712 M_1^7 \omega^6 M_0^2+\nonumber\\
&-&1443820371 M_2^3 M_1 M_0^5+1913406531 M_1^3 M_2^2 M_0^4+\nonumber\\
&-&17048199168 M_1^6 \omega^7 M_0^4+5499844064 M_1^6 \omega^3 M_0^4+\nonumber\\
&+&32426263584 M_1^7 \omega^4 M_0^2+5425958984 M_1^5 \omega^4 M_0^6+\nonumber\\
&+&1506160656 \omega M_2^4 M_0^4-18953494560 \omega^3 M_2^4 M_0^4+\nonumber\\
&+&57315834576 \omega^5 M_2^4 M_0^4+40359368 M_0^{10} M_1^3+\nonumber\\
&+&4325764872 M_1^4 M_2 \omega M_0^5-221128005384 M_2 M_1^5 \omega^4 M_0^3+\nonumber\\
&-&226503770769 M_2^3 \omega^4 M_1 M_0^5+49793449959 M_2^3 M_1 \omega^2 M_0^5+\nonumber\\
&-&9063952560 M_2^2 M_1^2 \omega M_0^6+273714152712 M_2 M_1^5 \omega^6 M_0^3+\nonumber\\
&+&237115503765 M_2^3 \omega^6 M_1 M_0^5+439217769789 M_2^2 M_1^3 \omega^4 M_0^4+\nonumber\\
&-&500779192221 M_2^2 \omega^6 M_1^3 M_0^4-84523620003 M_2^2 M_1^3 \omega^2 M_0^4+\nonumber\\
&+&11427580164 \omega M_2^2 M_1^4 M_0^2+75509594160 M_2 M_1^6 \omega^3 M_0+\nonumber\\
&-&10046350314 M_1^2 \omega M_2^3 M_0^3+269384667552 M_1^6 \omega^7 M_2 M_0+\nonumber\\
&+&38271033372 M_1^5 M_2 \omega^2 M_0^3+644064977556 M_1^4 \omega^5 M_2^2 M_0^2+\nonumber\\
&+&57847286016 M_2^2 M_1^2 \omega^3 M_0^6+146490916572 M_1^2 \omega^3 M_2^3 M_0^3+\nonumber\\
&-&140620399920 M_2^2 \omega^7 M_1^2 M_0^6+7441485480 M_2 M_1^4 \omega^5 M_0^5+\nonumber\\
&-&480543897834 M_1^2 \omega^5 M_2^3 M_0^3+418419689688 M_1^2 \omega^7 M_2^3 M_0^3+\nonumber\\
&-&590669679600 M_1^4 \omega^7 M_2^2 M_0^2-183306795672 M_1^4 \omega^3 M_2^2 M_0^2+\nonumber\\
&+&4067972064 M_2^2 M_1^2 \omega^5 M_0^6-279791936424 M_1^6 \omega^5 M_2 M_0+\nonumber\\
&-&36119871216 M_2 M_1^4 \omega^3 M_0^5)
\label{ges}
\end{eqnarray}

\section{Acknowledgments}
This  work  was partially supported by the Spanish  Ministry of Education and Science
 under Research Project No. FIS 2006-05319, and the Consejer\'\i a de Educaci\'on of the Junta de Castilla y 
Le\'on under the Research Project Grupo de Excelencia GR234.

\end{document}